\documentclass[useAMS,usenatbib]{mn2e}
\usepackage{graphicx}
\usepackage{lscape}
\usepackage{rotating}
\usepackage{longtable}
\newcommand{\kms}{\mbox{km\,s$^{-1}$}}
\newcommand{\teff}{$T_{\rm eff}$}

\title[High resolution abundances of the Hyades Supercluster]{High resolution elemental abundance analysis of the Hyades Supercluster \thanks{Based on observations obtained at the European Southern Observatory, Paranal, Chile (ESO programmes 080.D-0094(A) and 381.B-0045(A)).}}
\author[De Silva, Freeman, Bland-Hawthorn et~al.]  {G.M. De Silva$^{1}$\thanks{E-mail: gdesilva@aao.gov.au}, K.C. Freeman$^{2}$, J. Bland-Hawthorn$^{3}$, M. Asplund$^{4}$, M. Williams$^{5}$,  \newauthor J. Holmberg$^{6}$\\\\
$^{1}$Australian Astronomical Observatory, PO Box 296, NSW 1710, Australia\\
$^{2}$Research School of Astronomy and Astrophysics, Mount Stromlo Observatory, Australian National University, ACT 2611, Australia\\
$^{3}$Institute of Astronomy, School of Physics, University of Sydney, NSW 2006, Australia\\
$^{4}$Max Planck Institute for Astrophysics, Karl-Schwarzschild-strasse 1, Garching, 85741 Germany\\
$^{5}$Astrophysikalisches Institut Potsdam, An der Sternwarte 16, Potsdam D-14482, Germany\\
$^{6}$The Niels Bohr Institute, Astronomy Group, Juliane Maries Vej 30, 2100 Copenhagen, Denmark\\
}

\begin{document}

\date{Accepted. Received ; in original form }

\pagerange{\pageref{firstpage}--\pageref{lastpage}} \pubyear{2011}

\maketitle

\label{firstpage}

\begin{abstract}
The existence of a kinematically defined moving group of stars centred at U = -40, V = -17 \kms\ referred to as the Hyades Supercluster, has been suggested as the debris of an originally large star forming event, with its core being the present day Hyades open cluster. Using high-resolution UVES spectra, we present elemental abundances for a range of alpha, Fe-peak and neutron-capture elements for 26 proposed supercluster stars. Our results show that the sample stars display a heterogeneous abundance distribution, with a clump around [Fe/H] = +0.15.  We also calculate stellar radial velocities and U,V,W space velocities. Enforcing a strict chemical and kinematical membership criteria, we find 4 supercluster stars share the Hyades open cluster abundances and kinematics, while many of the remaining stars fit the disc field kinematics and abundance range. We discuss our findings in the context of the Hyades supercluster being a dispersed star-forming remnant, a stellar stream of purely dynamical origin or a result of several processes.
\end{abstract}


\section{Introduction}

\label{lastpage}
To obtain a detailed physical understanding of the events that shaped the Galactic disc, we must examine its individual stars. The chemical abundances as measured from stellar photospheres can be indicative of the conditions of a star's birth site. Different chemical elements are synthesized during different stages of stellar evolution and distributed to the interstellar medium via supernova explosions or stellar winds. By studying the abundances of these elements we can estimate the frequency and importance of each synthesis process. Therefore the chemical abundance patterns within the disc stars can be regarded as a fossil imprint and hold key information of the events that took place during the disc's formation and evolutionary history.  The long term goal of Galactic Archaeology is to reconstruct the original star-forming events of the Galaxy. Employing the technique of chemical tagging \citep{fbh02} to disc stars, one can identify now dispersed ancient stellar aggregates, the building blocks of the Galactic disc. \\

Candidates for such fossil stellar aggregates are moving groups and superclusters, where their stellar members are unbound but share a common motion around the Galaxy. Analogous to stellar streams in the Galaxy halo which are remnants of merged satellites, old moving groups and superclusters may represent the ancient stellar building blocks of the disc. A strong advocate for the reality of these groups was Olin Eggen, who identified several moving groups and superclusters (Eggen 1958 -- 1998). With the availability of the Hipparcos catalogue of parallaxes and proper motions it became possible to study these substructures in unprecedented detail. Several studies have analysed the Hipparcos catalogue and re-identified the classical groups as well as found new ones \citep{antoja,zhao,bovy09}. Essentially, all of these studies have focused purely on the dynamical/kinematical nature of the substructures in the disc.\\

Thus moving groups and superclusters provide disc substructures that can be identified through kinematics. However, it is not yet clear if all, or indeed the majority,
of these substructures actually originate from star forming clusters that have or are in the process of being dispersed. Such substructures can equally well be the result of the dynamical interaction between the stars in the field and the central bar  or the spiral structure of the Milky Way \citep[see e.g.][] {dehnen98}. Therefore not all kinematically defined disc substructures represent the debris of a single star formation event. 

\subsection{The Hyades Supercluster}
The existence of a dispersed kinematically defined group of stars associated with the Hyades open cluster was proposed by \citet{e70}, as part of a larger star-forming event, where the outer cluster stars have dispersed into the Hyades supercluster, leaving only its core members in the present day Hyades open cluster. The existence of the kinematically defined Hyades supercluster is well established observationally (e.g. in Hipparcos, RAVE and Geneva-Copenhagen survey), where a concentration of stars are found at U $\sim$ -40 \kms\ and V $\sim$ -20 \kms\ in the $U-V$ velocity plane. Throughout this article we adopt the direction of the U-velocity as positive towards the Galactic centre. However, many studies suggest that this over density of stars consists of a mixture of field stars that have been resonantly swept up by a passing spiral wave and is not the results of a dispersed stellar aggregate \citep{famaey, minchev, bovy10}. \\

With kinematical information alone it is difficult to address the reality of dispersed stellar aggregates. The dissipative formation and dynamical evolution of the Galaxy means much of the kinematical information is limited to the last event of scattering, where the original kinematical information has been modified as a result of latter events. Therefore only the most recent events can be traced with kinematics, masking the initial identity. However the chemical information in the form of elemental abundances of individual stars remains preserved. Present day open clusters are chemically homogeneous \citep{desilva06,pancino2010}, which demonstrates that the stars natal chemical composition is unaltered. Therefore, along with kinematical information, chemical probing of moving groups and superclusters offers the possibility to determine their initial identity. Detailed abundance study of the HR1614 moving group \citep{hr1614}  demonstrated its member stars are chemically homogeneous and clearly distinguishable from the disc field stars, indicating that it is a dispersed relic of an earlier star-forming event \citep[see also][for the case of Wolf 630 moving group]{BubarKing}. Conversely, an example of a dynamically defined group is the Hercules stream, which is not chemically homogeneous and its abundance patterns are not distinguishable from the field stars \citep{hercules}. When inspected in velocity space alone, both the HR1614 moving group and the Hercules stream satisfy the requirements for a dispersing cluster remnant, while chemical probing shows that they have very different origins. These examples demonstrate that chemical information is essential to identify dispersed group members, and that dynamical information alone cannot be used to uncover the true history behind any co-moving group of stars in the disc.\\

Despite the wealth of information obtainable via chemical probing, elemental abundance analysis of the kinematically defined moving groups are lacking in the literature, possibly due to the larger effort required to measure accurate abundances. In this paper we present high resolution elemental abundances for a sample of proposed Hyades supercluster stars and examine its possible origins.

\section{Sample selection}

Assuming that the Hyades supercluster is part of larger structure produced by a dissolving star formation site, we can expect that the site would dissolve along a dispersion orbit. These paths along which a gently disrupting system would dissolve, can be approximated by closed ellipses when viewed from a frame rotating at the local value of $\Omega$ - $\kappa/2$ where $\Omega$ is the local angular velocity of rotation near the sun and $\kappa$ is the epicyclic frequency. The stars move clockwise around these ellipses with angular velocity approximately $\kappa/2$ and gradually spread out around them. In this work, the adopted velocity of Hyades supercluster stars near the sun was U = -40, V = -17 \kms\, consistent with the currently accepted motion of the Hyades open cluster and the corresponding over-density regions seen in the Geneva-Copenhagen survey.\\
  
\citet{wilson} observed low latitude K giant stars from the Michigan catalogue \citep{Houk} in the expected direction of the Hyades supercluster dispersion orbit. The observations were carried out using the coude spectrograph on the Mt Stromlo Observatory 74inch telescope. The data of R $\sim$ 20,000 centred at 5200$\AA\ $ were cross-correlated against template spectra to derive metallicities and radial velocities. Wilson found a concentration of
stars in the abundance range -0.2 to -0.1 lying close to the Hyades dispersion orbit loci in the radial velocity - longitude plane, which he interpreted as being stars associated with the dispersing Hyades systems. The currently accepted Hyades open cluster metallicity is [Fe/H] = +0.13 dex based on high resolution abundances \citep{paulson,primas}. While we cannot accurately compare high resolution Fe abundances with the early metallicity derivations due to likely zero-point offsets, it is interesting that a concentration of stars was found along the dispersion orbits in this metallicity bin. \\

We selected stars likely to be dispersing Hyades supercluster stars based on Wilson's results for high resolution observations. Our analysis will test the hypothesis of the disrupting system discussed above. We also targeted F-type probable member stars as published in \citet{e98}, which had Hipparcos identification. Table \ref{sample} lists the IDs, co-ordinates, V magnitude and colour for all observed stars. In the last column of Table \ref{sample}, E98 indicates stars selected from \citet{e98} and W indicates those selected from Wilson's thesis. \\ 

\begin{table*}
\caption{Total sample of observed stars}             % title of Table
\label{sample}      % is used to refer this table in the text
\centering                          % used for centering table
\begin{tabular}{c c c c c c}        % centered columns (4 columns)
\hline\hline                 % inserts double horizontal lines
ID & RA & DEC & V & B-V & notes\\    % table heading 
\hline    
HD 25102 & 03 59 40.4935 & +10 19 49.448 & 6.34 & 0.42 & E98 \\
HD 26737 & 04 14 30.4203 & +22 27 06.713 & 7.04 & 0.43 & E98 \\
HD 27429 & 04 20 25.1095 & +18 44 33.390 & 6.09 & 0.38 & E98 \\
HD 27534 & 04 21 32.2686 & +18 25 03.297 & 6.78 & 0.45 & E98 \\
HD 27848 & 04 24 22.2727 & +17 04 44.225 & 6.95 & 0.46 & E98 \\ 
HD 27991 & 04 25 37.3171 & +15 56 27.636 & 6.45 & 9.45 & E98 \\
HD 28394 & 04 29 20.5531 & +17 32 41.766 & 7.01 & 0.51 & E98 \\
HD 28608 & 04 30 57.1723 & +10 45 06.365 & 7.02 & 0.47 & E98 \\
HD 28736 & 04 32 04.8091 & +05 24 36.126 & 6.35 & 0.43 & E98 \\
HD 30912 & 04 52 47.1170 & +27 53 50.947 & 5.97 & 0.34 & E98 \\
HD 50643 & 06 53 21.8408 & -18 55 58.238 & 6.13 & 0.16 & E98 \\
HD 69511 & 08 15 58.8226 & -35 54 11.486&  6.16 & 1.57 &W\\
HD 69836 & 08 17 30.5546 & -35 10 14.009&  8.57 & 1.08 &W\\ 
HD 72320 & 08 30 29.1090 & -42 54 34.058&  8.95 & 0.94 &W\\ 
HD 72630 & 08 32 23.7142 & -41 29 53.544&  9.49 & 1.01 &W\\ 
HD 73657 & 08 37 48.1209 & -42 19 11.404&  7.85 & 1.18 &W\\ 
HD 73829 & 08 38 44.3768 & -43 08 16.205&  9.42 & 1.27 &W\\ 
HD 74165 & 08 40 53.4114 & -40 40 55.841&  9.12 & 1.16 &W\\ 
HD 74166 & 08 40 42.6510 & -42 20 26.606&  7.62 & 1.29 &W\\ 
HD 74529 & 08 42 43.4206 & -47 14 25.097&  8.41 & 1.19 &W\\ 
HD 74900 & 08 45 13.1890 & -41 52 44.224&  7.96 & 1.17 &W\\ 
HD 75058 & 08 46 21.9930 & -38 18 19.552&  8.94 & 1.08 &W\\ 
HD 75171 & 08 44 29.9575 & -65 49 31.544 & 6.03 & 0.19 & E98 \\ 
HD 76128 & 08 52 47.7716 & -41 25 49.269&  7.90 & 1.40 &W\\ 
HD 77117 & 08 58 37.7029 & -47 04 17.465&  9.46 & 0.52 &W\\
HD 77241 & 08 59 30.6086 & -48 49 11.118&  9.98 & 1.02 &W\\ 
HD 78002 & 09 04 19.8280 & -43 20 22.225&  9.45 & 1.05 &W\\ 
HD 78097 & 09 04 46.8967 & -44 48 37.337&  7.61 & 1.63 &W\\ 
HD 78204 & 09 05 02.8562 & -50 52 45.128&  9.95 & 1.04 &W\\ 
HD 78402 & 09 06 16.1539 & -49 04 41.490&  9.88 & 0.93 &W\\ 
HD 78528 & 09 06 57.8364 & -49 38 22.415&  8.86 & 1.59 &W\\ 
HD 78959 & 09 09 29.8423 & -44 33 59.972&  7.80 & 1.75 &W\\ 
HD 80571 & 09 19 07.6889 & -46 38 22.151&  7.88 & 1.09 &W\\ 
HD 80911 & 09 20 41.4133 & -50 42 09.173&  8.56 & 1.73 &W\\ 
HD 81278 & 09 22 52.1556 & -52 08 04.264&  8.48 & 1.08 &W\\ 
HD 82438 & 09 30 12.8630 & -52 01 55.341&  9.00 & 1.64 &W\\ 
HD 83234 & 09 35 04.8294 & -54 36 14.833&  8.56 & 1.42 &W\\ 
HD 84598 & 09 44 53.2380 & -50 00 36.850&  7.45 & 0.97 &W\\ 
HD 86757 & 09 58 56.8698 & -54 37 13.983&  8.36 & 1.59 &W\\
HD 97840 & 11 15 02.4004 & -33 19 03.880 & 6.99 & 0.32 & E98 \\
HD 112734 &12 58 33.8230 & +28 19 10.449 & 6.94 & 0.26 & E98 \\
HD 117374 &13 40 55.4795 & -85 47 09.756 & 5.56 & 0.17 & E98 \\
HD 121164 &13 53 10.2822 & +28 38 53.273 & 5.91 & 0.19 & E98 \\
HD 122721 &14 04 14.5737 & -33 06 19.948 & 9.16 & 1.06 & E98 \\
HD 218475 &23 08 22.9309 & -07 16 01.221 & 8.31 & 0.33 & E98 \\
\hline                                   %inserts single line
\end{tabular}
\end{table*}

\section{Observations}
High resolution and high signal-to-noise spectra of probable members of the Hyades supercluster were observed using VLT UV-Visual Echelle Spectrograph (UVES) at UT2 in the framework of programs 080.D-0094(A) and 381.B-0045(A). A total of 45 probable member stars were submitted for service mode observations, using the UVES Red arm standard setting at 520nm which provides complete spectral coverage from 4200\AA\ to 6200\AA, and employed a 0.8arcsec slit to achieve a spectral resolving power of 60,000. The typical S/N ratio was 100 per pixel. The data were reduced with the latest UVES ESO-MIDAS pipeline. The resulting spectra were normalized using the $continuum$ task in the IRAF\footnote{IRAF is distributed by the National Optical Astronomy Observatory, which is operated by the Association of Universities for Research in Astronomy, Inc., under cooperative agreement with the National Science Foundation.} package. \\

Out of the total 45 targets observed, only 26 could be used for abundance analysis. The other 19 stars showed broadened spectral lines due to high rotation and were hence discarded from further analysis. 17 of these stars were from \citet{e98}, likely to be F-type dwarfs, and two stars were K giants from Wilson's thesis. Of the 26 sharp-lined spectra, only one star was from the sample of \citet{e98}. \\

\section{Abundance analysis}

\subsection{Model Atmospheres and Spectral Lines}
The elemental abundances were derived based on EW measurements and spectral synthesis, making use of the latest version of the MOOG code \citep{sneden73}. The EWs were measured by fitting a Gaussian profile to each line using the interactive $SPLOT$ function in IRAF. On occasions the local continuum level was adjusted by eye. Abundances of Mn, Ba, La and Ce were derived using the $synth$ routine of the MOOG code to account for line blending and hyperfine structures. Interpolated Kurucz model atmospheres based on the ATLAS9 code \citep{Castelli97} with no convective overshoot were used throughout this study.\\

The line list used in this analysis is available in the online version. The employed lines and $gf$ values for Fe, Na, Mg, Al, Si, Ca, Ti, Cr, Ni, and Zn were a subset of the lines used by \citet{bensby03}, where their main source of the Fe~{\sc i} line data is the laboratory measurements by the Oxford group \citep[and references therein]{blackwell79a, blackwell79b, blackwell95}. We chose this particular Fe line list as it will better enable comparison of our results with those of the disc field stars by \citet{bensby03,bensby05}, as well as with the Hyades open cluster abundances of \citet{primas} who employ the same line list. The lines of S, Sc, Co, and Zr were adopted from \citet{gratton01}.  For Mn and Ba, the line data were taken from \citet{prochaskamcwill} and \citet{prochaska00} respectively and include the effects of hyperfine splitting. The La and Ce line data were obtained via the VALD database \footnote{http://ams.astro.univie.ac.at/vald/} \citep{VALD,VALD1,VALD2,VALD3}.\\

\subsection{Stellar parameters and Elemental abundances}

\begin{figure}
  \centering
  \includegraphics[scale=0.32]{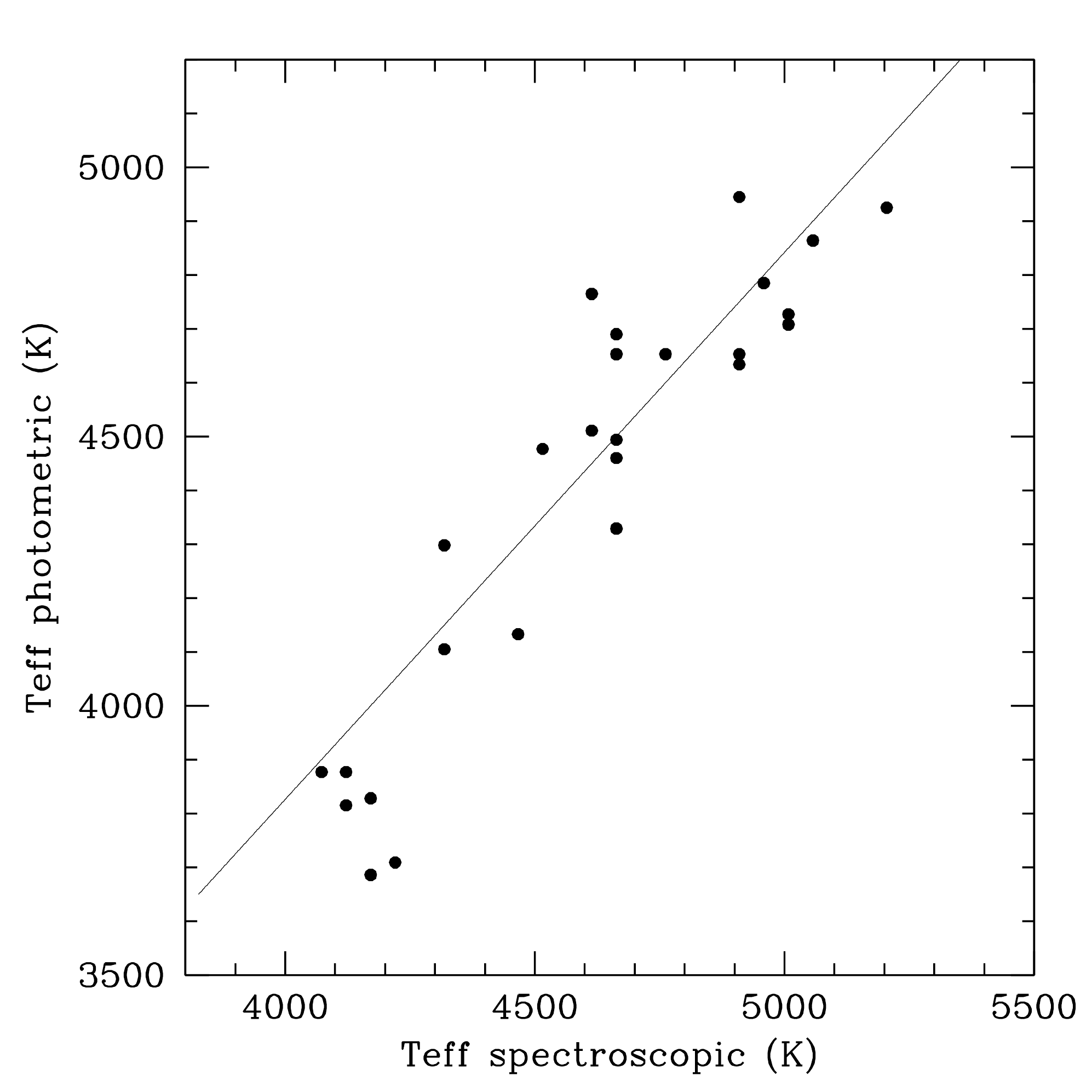}
     \caption{Effective temperatures via the $B-V$ colour calibrations of Alonso et al 1999 vs. the spectroscopic temperatures. The solid line is the average difference, where the spectroscopic values are hotter on average by $\sim$150K.}
        \label{color_teff}
  \end{figure}

We derive the stellar parameters based on spectroscopy. Abundances for all Fe {\sc i} and {\sc ii} lines were computed from the measured EWs. Effective temperature (\teff) was derived by requiring excitation equilibrium of the Fe~{\sc i} lines. Microturbulence was derived from the condition that Fe {\sc i} lines show no trend with EW. Surface gravity (log g) was derived via ionization equilibrium, i.e. requiring the abundances from Fe {\sc i} equals Fe {\sc ii}.  The resulting stellar parameters are given in Table \ref{params}. For comparison we also calculated stellar temperatures based on photometry. Using the $B-V$ colour value and adopting [Fe/H] = 0.0 we use equation (4) of \citet{alonso99} to derive photometric stellar temperatures. We find that the spectroscopic effective temperatures are on average, hotter by approx 150K compared to the photometric values. Fig \ref{color_teff} plots the photometric effective temperature against the spectroscopic values.\\

\begin{figure}
  \centering
  \includegraphics[scale=0.32]{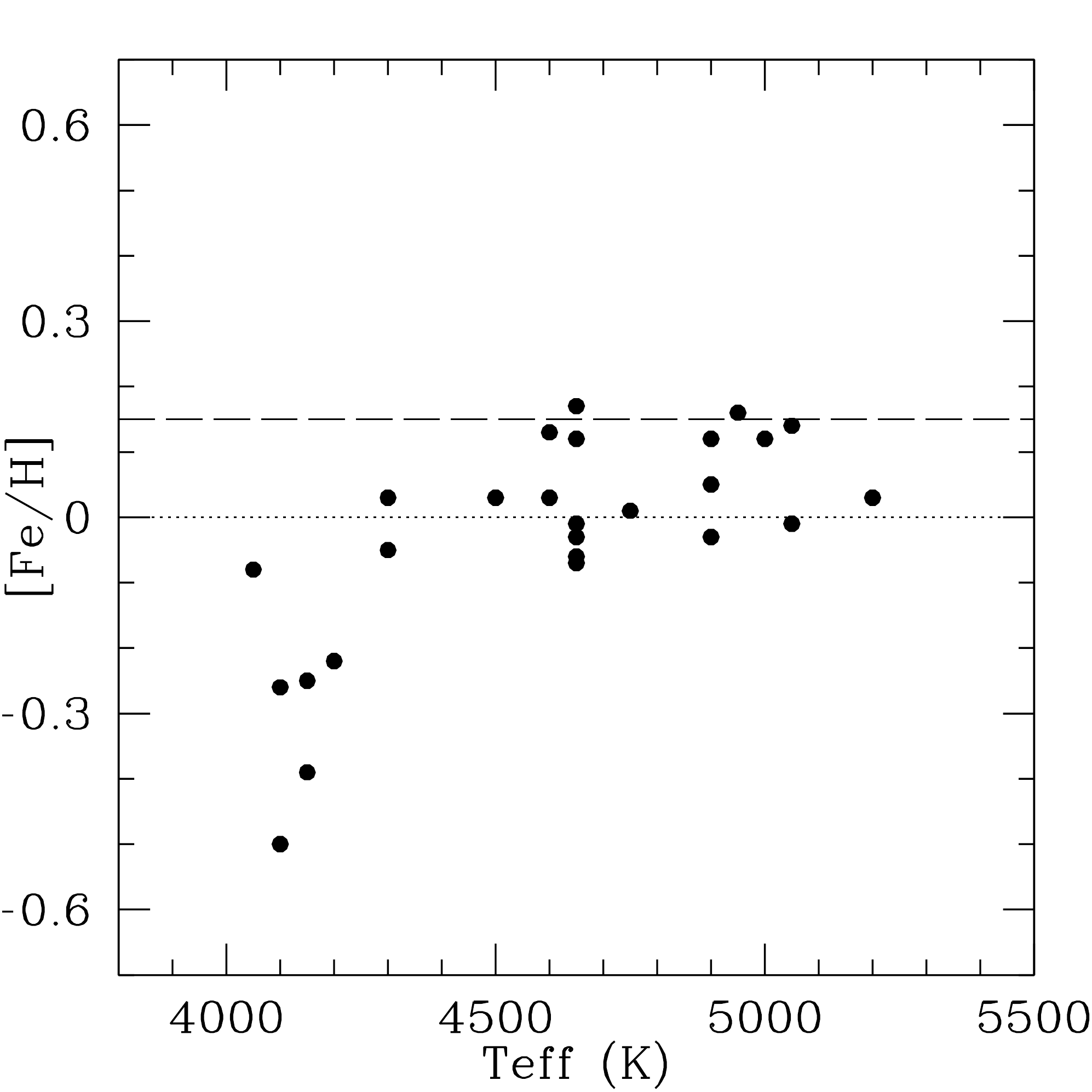}
     \caption{[Fe/H] vs. effective temperature. The dotted line represented the solar level and the dashed line represents the Hyades open cluster metallicity.}
        \label{fe_teff}
  \end{figure}

To check for possible trends with effective temperature, we plot the spectroscopic [Fe/H] vs. the spectroscopic effective temperature in Fig \ref{fe_teff}. The dotted line represents the solar metallicity \citep[adopted from][]{asplund09} and the dashed line represents the Hyades open cluster metallicity of [Fe/H] = 0.15 (Note that \citet{paulson} finds the Hyades open cluster to have [Fe/H] = 0.13, but taking into account the difference in the solar metallicity used in this study, we place the Hyades open cluster at [Fe/H] = 0.15). The six coolest stars show a sudden drop in metallicity. As there is no clear Fe abundance trends with \teff, it is unlikely that we are seeing a temperature dependent effect or non-LTE effects.
\\

\begin{table*}
\caption{Spectroscopic stellar parameters}             % title of Table
\label{params}      % is used to refer this table in the text
\centering                          % used for centering table
\begin{tabular}{c c c c c c}        % centered columns (4 columns)
\hline\hline                 % inserts double horizontal lines
Star ID & T$_{eff}$(K) & log $g$(cm~s$^{-2}$) & $\xi$(\kms) \\    % table heading 
\hline    
HD86757 &  4100& 0.5& 2.3  \\
HD84598 &  5050& 2.8& 1.45 \\
HD83234 &  4300& 2.0& 1.6  \\
HD82438 &  4100& 0.5& 2.6  \\
HD81278 &  4900& 3.0& 1.1  \\
HD80911&  4200& 0.7& 2.3  \\
HD80571 &  4900& 2.5& 1.4 \\
HD78959 &  4150& 1.2& 2.2 \\
HD78528 & 4050& 1.2& 1.7  \\
HD78402 & 4900& 2.5& 1.1  \\
HD78204 & 5050& 2.8& 1.5  \\
HD78097 & 4150& 0.6& 2.4  \\
HD78002 & 5000& 3.0& 1.4  \\
HD77241 & 4600& 2.3& 1.6  \\
HD76128 & 4650& 2.8& 1.3  \\
HD75058 & 4650& 2.8& 1.2  \\
HD74900&   4650& 2.6& 1.5 \\
HD74529& 4650& 2.6& 1.4 \\
HD74166&   4300& 1.8& 1.8\\ 
HD74165&  4600& 2.4& 1.5  \\
HD73829&  4650& 2.5& 1.45 \\
HD73657&   4500& 2.2& 1.7 \\
HD72630 & 4950& 3.0& 1.4  \\
HD72320&   5200& 3.2& 1.4 \\
HD69836&   4750& 2.5& 1.45\\
HD122721& 4650& 2.5& 1.25 \\
\hline                                   %inserts single line
\end{tabular}
\end{table*}

The reason for the large drop in [Fe/H] for the cool stars is unclear. We re-computed the Fe abundances for these cool stars using the photometric \teff\ values. We find this results in Fe {\sc i} abundances increasing closer to the warmer star abundance levels, however large deviations of the order of +0.2dex are seen between Fe {\sc i} and Fe {\sc ii} abundances, where Fe {\sc ii} is larger. Studies of cool K-dwarfs in the Hyades open clusters and in the disc field have reported similar discrepancies between Fe {\sc i} and Fe {\sc ii} and other ionized elements when adopting photometric temperature scales \citep{yong04, s4n}. Chromospheric activities associated with magnetic fields are thought to give rise to these effects in the cool dwarfs. However our targets are K-giants, and we are not aware of any studies indicating such discrepancies among cool giants. Given the large Fe I and Fe II discrepancies when using the photometric temperatures, and the significantly lower metallicities resulting from the spectroscopic parameters, we cannot rely on either set of parameters for the cool stars. We hope to further investigate this issue in a dedicated study.  As this is beyond the aims and scope of the present context, we will refrain from considering the six coolest stars from further analysis in this article.\\

\begin{figure}
  \centering
  \includegraphics[scale=0.4]{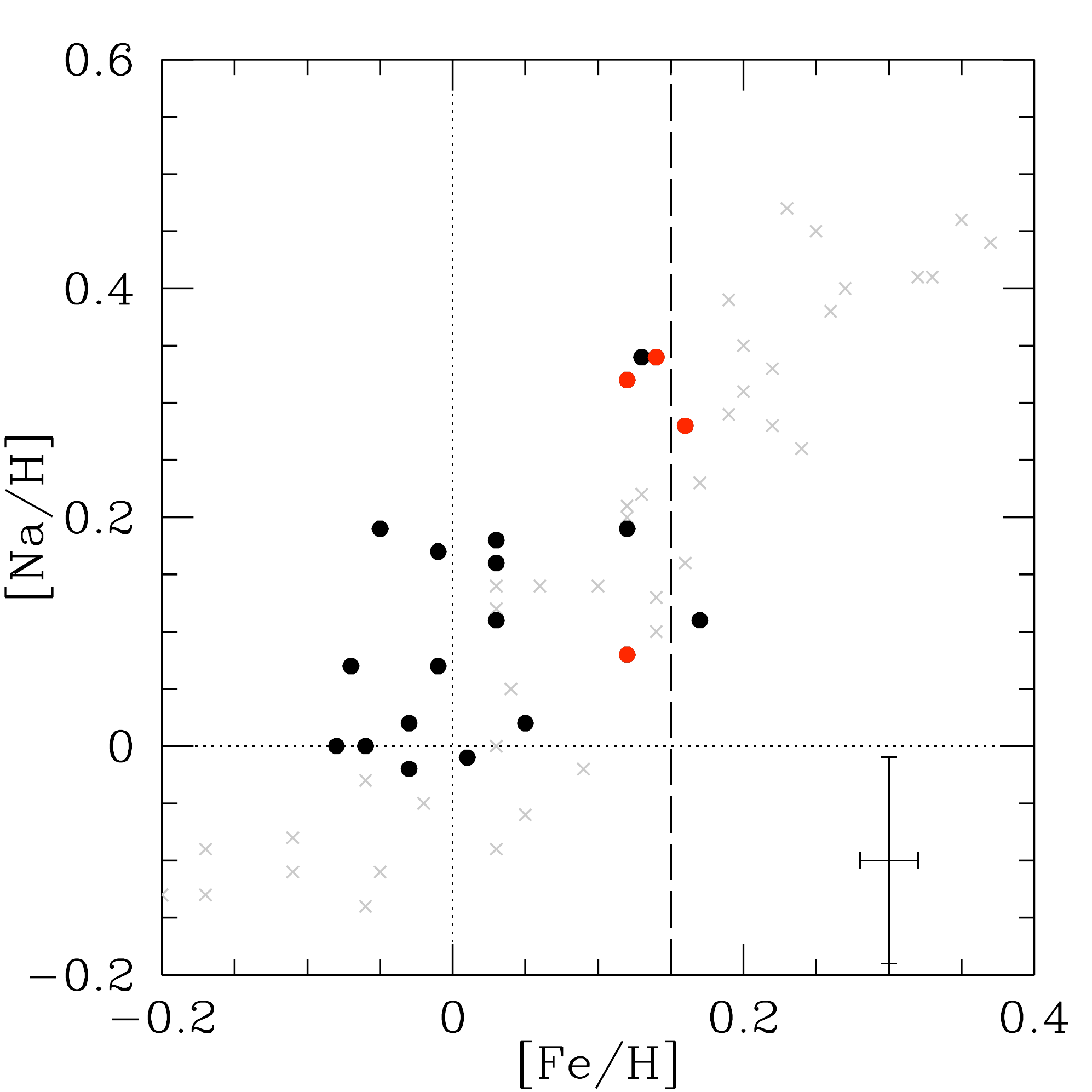}
  \includegraphics[scale=0.4]{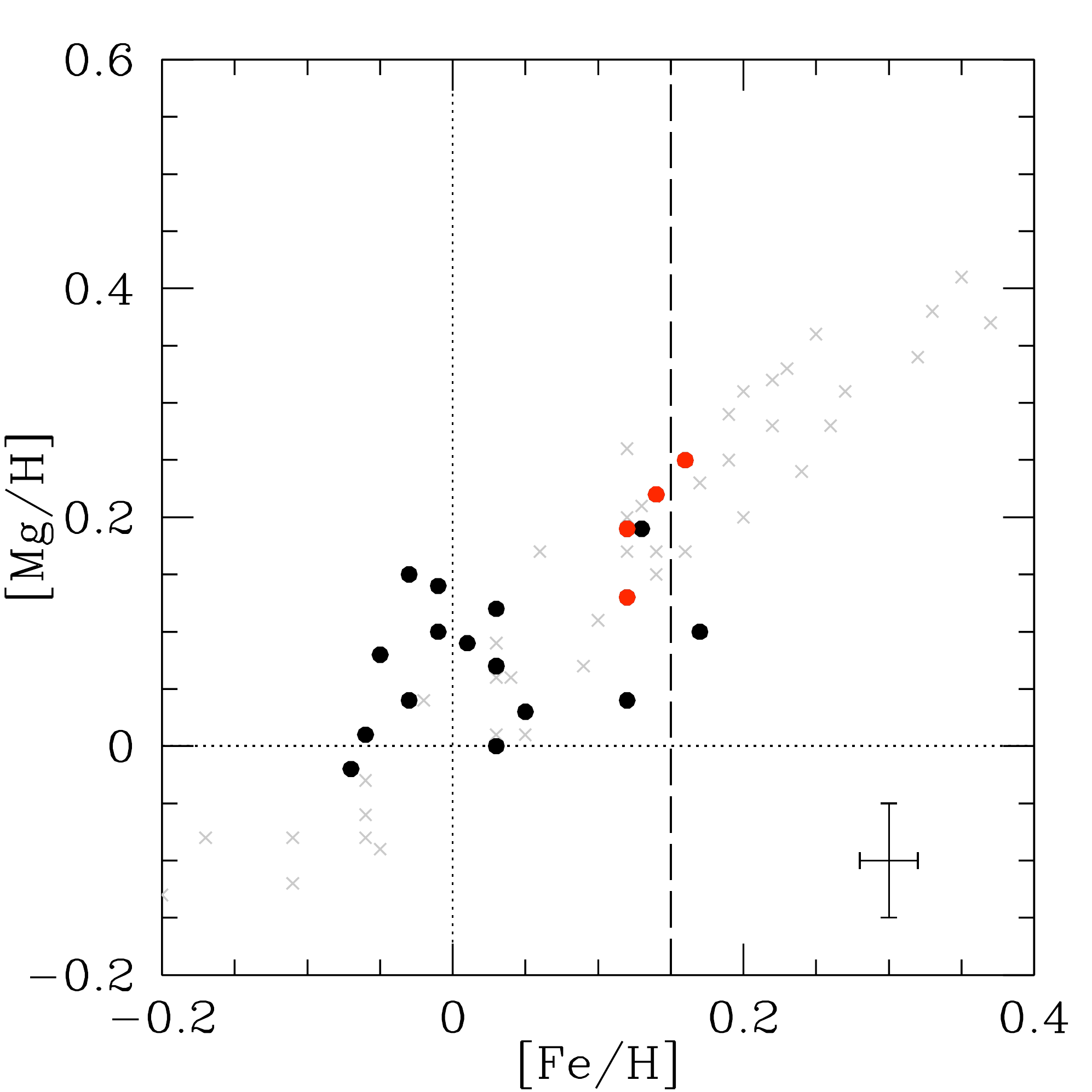}
     \caption{[X/H] vs. [Fe/H] for the sample stars (filled circles). The grey crosses are thin disc stars of \citet{bensby05}.}
        \label{abplot_namg}
  \end{figure}

\begin{figure}
  \centering
  \includegraphics[scale=0.4]{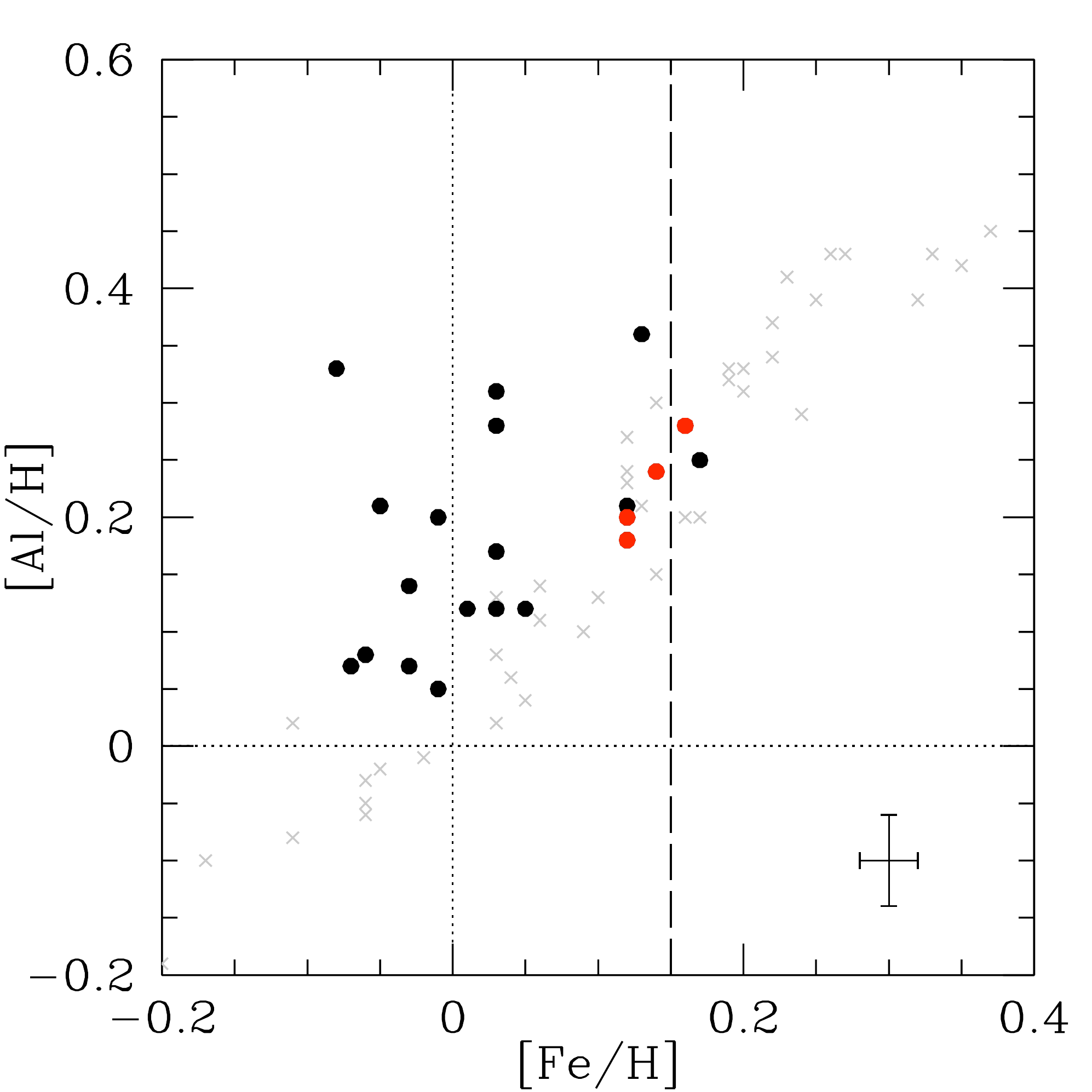}
  \includegraphics[scale=0.4]{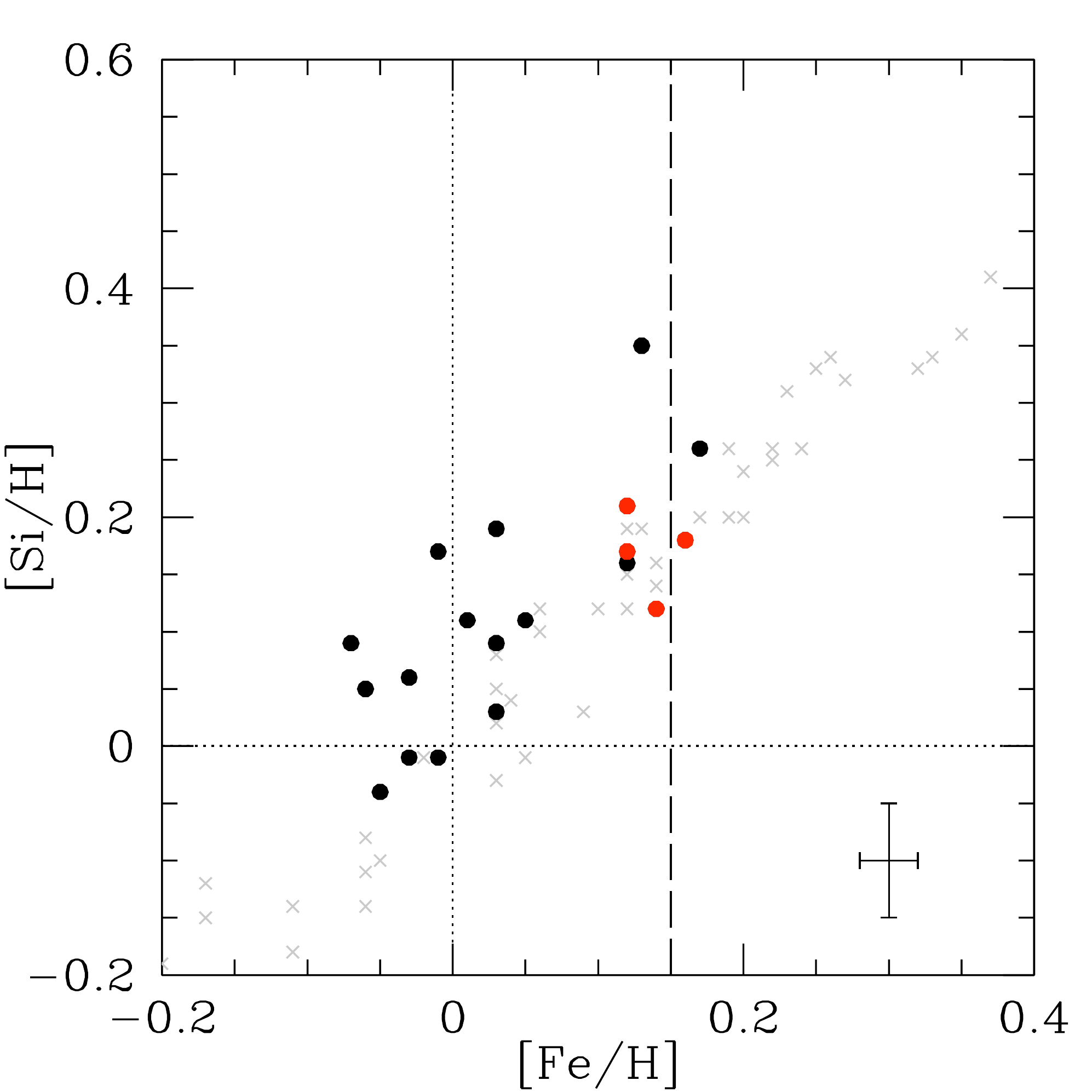}
     \caption{[X/H] vs. [Fe/H] for the sample stars (filled circles). The grey crosses are thin disc stars of \citet{bensby05}.}
        \label{abplot_alsi}
  \end{figure}

\begin{figure}
  \centering
  \includegraphics[scale=0.4]{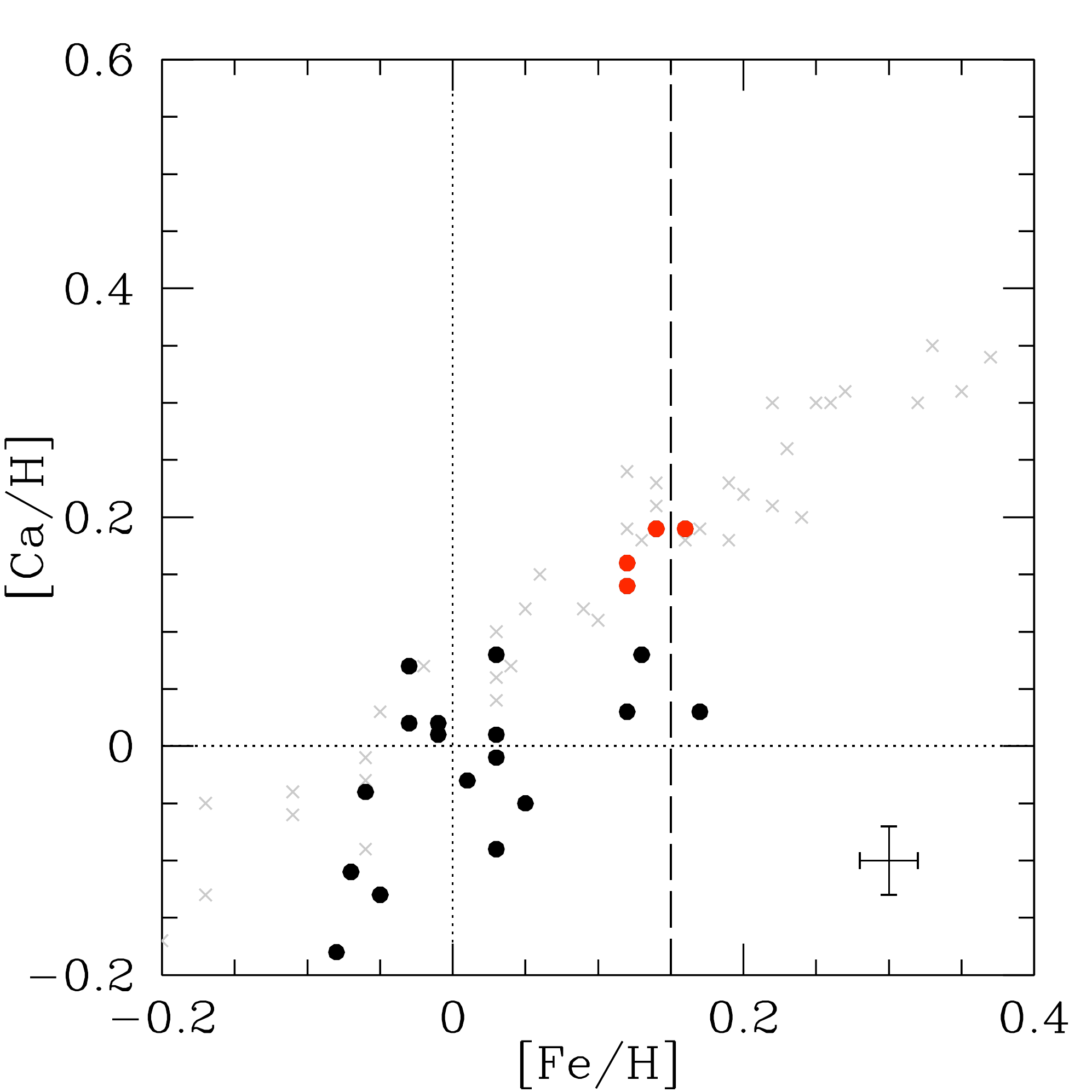}
  \includegraphics[scale=0.4]{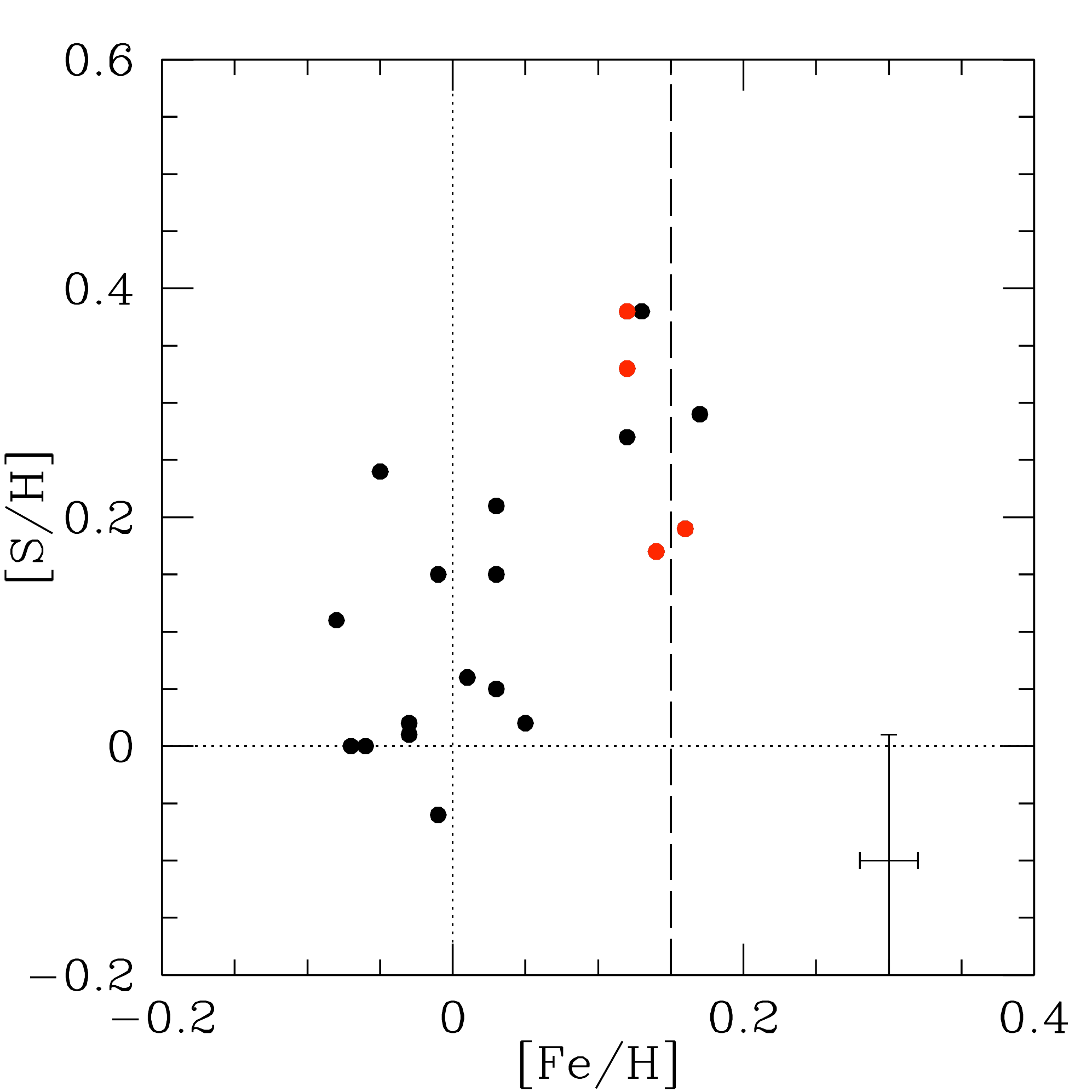}
     \caption{[X/H] vs. [Fe/H] for the sample stars (filled circles). The grey crosses are thin disc stars of \citet{bensby05}.}
        \label{abplot_sca}
  \end{figure}

\begin{figure}
  \centering   
  \includegraphics[scale=0.4]{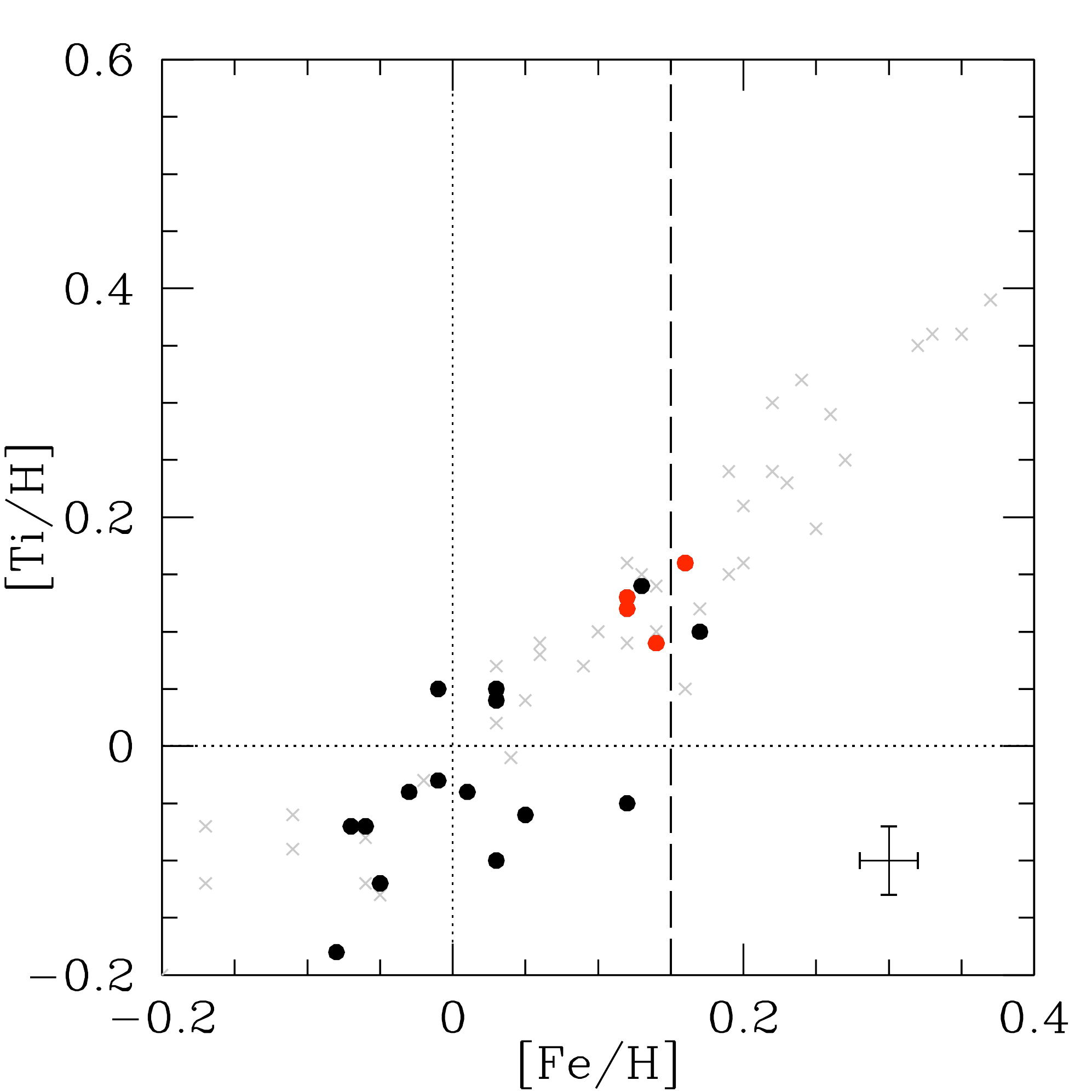}
  \includegraphics[scale=0.4]{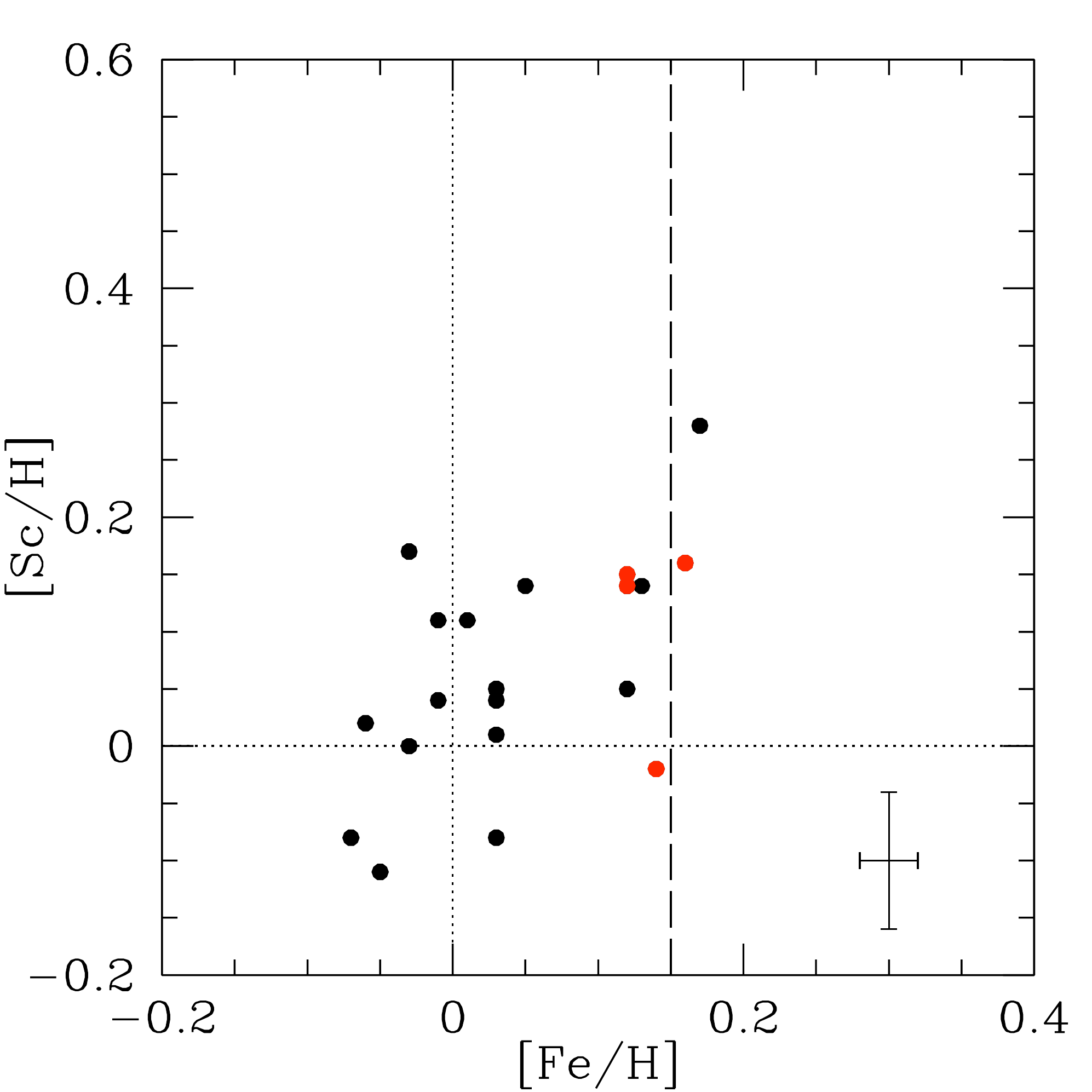}
     \caption{[X/H] vs. [Fe/H] for the sample stars (filled circles). The grey crosses are thin disc stars of \citet{bensby05}.}
        \label{abplot_scti}
  \end{figure}

\begin{figure}
  \centering
  \includegraphics[scale=0.4]{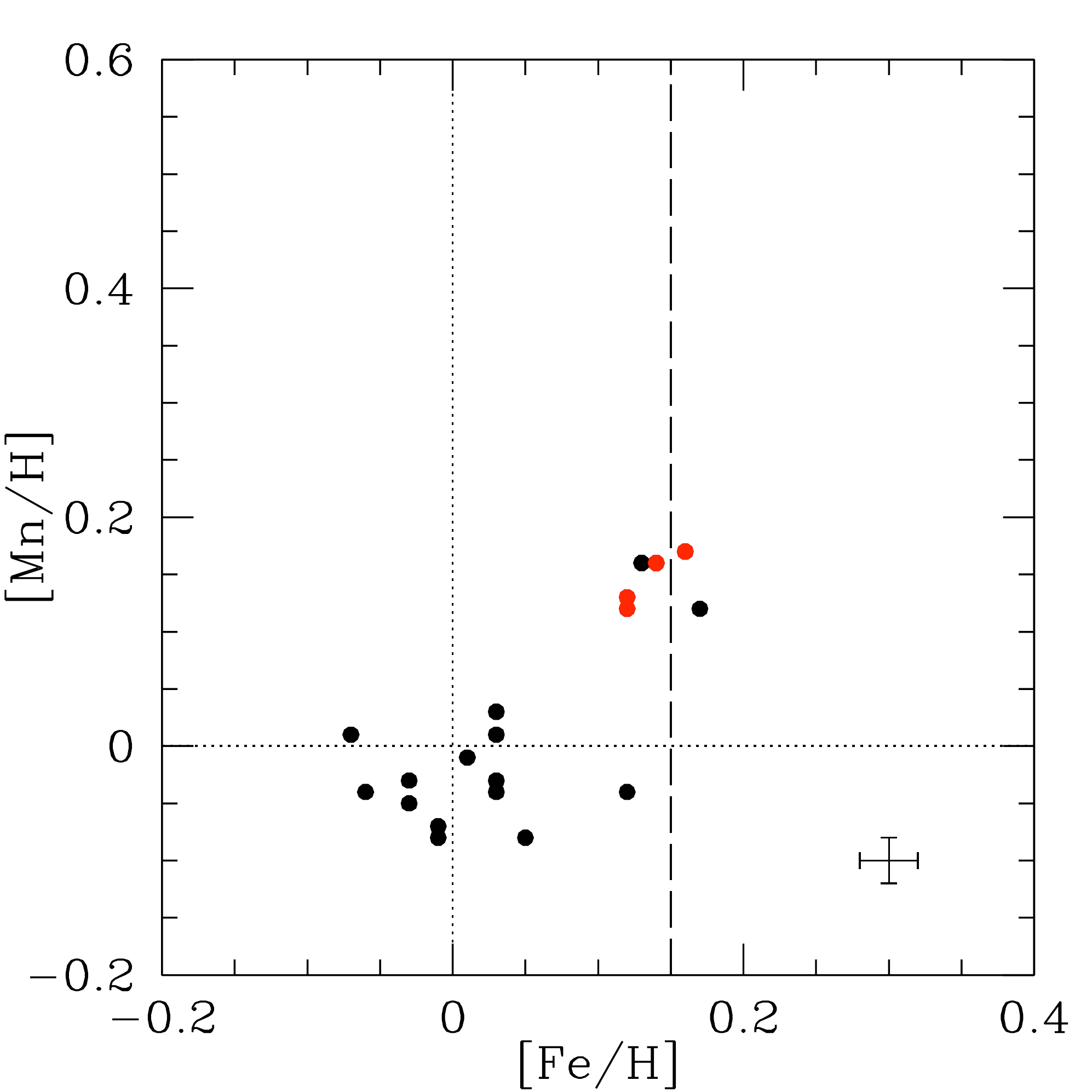}
  \includegraphics[scale=0.4]{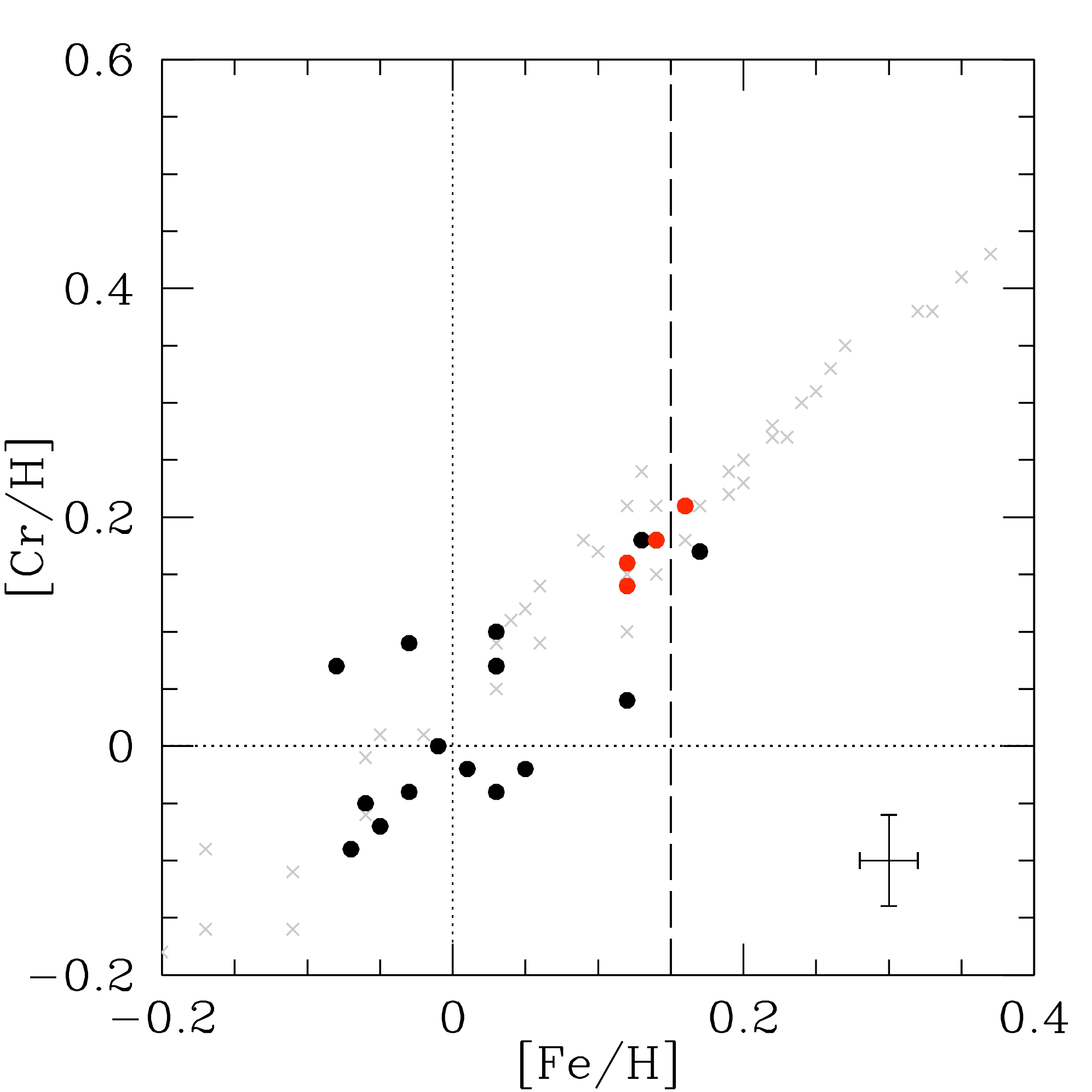}
     \caption{[X/H] vs. [Fe/H] for the sample stars (filled circles). The grey crosses are thin disc stars of \citet{bensby05}.}
        \label{abplot_crmn}
  \end{figure}

\begin{figure}
  \centering
  \includegraphics[scale=0.4]{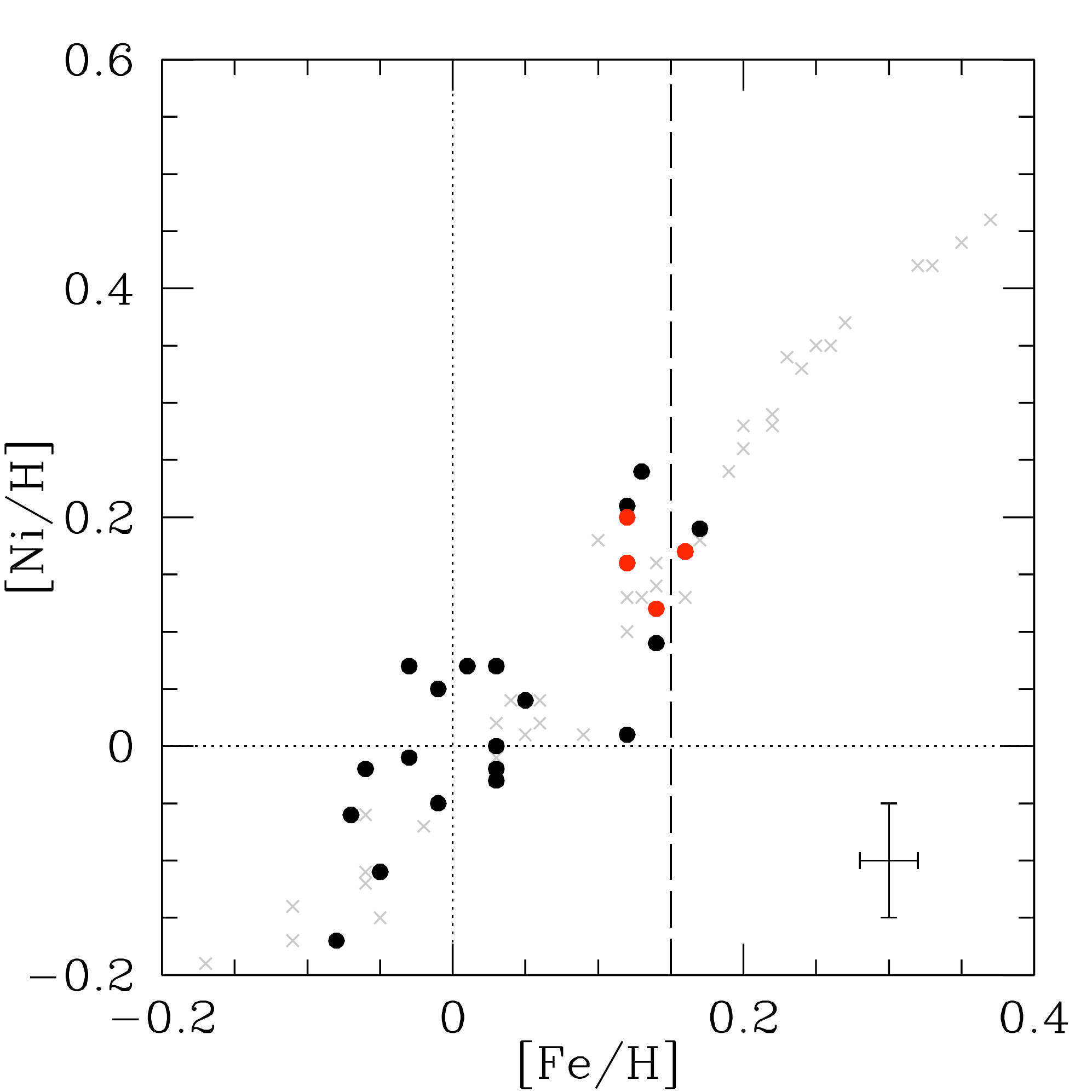}
  \includegraphics[scale=0.4]{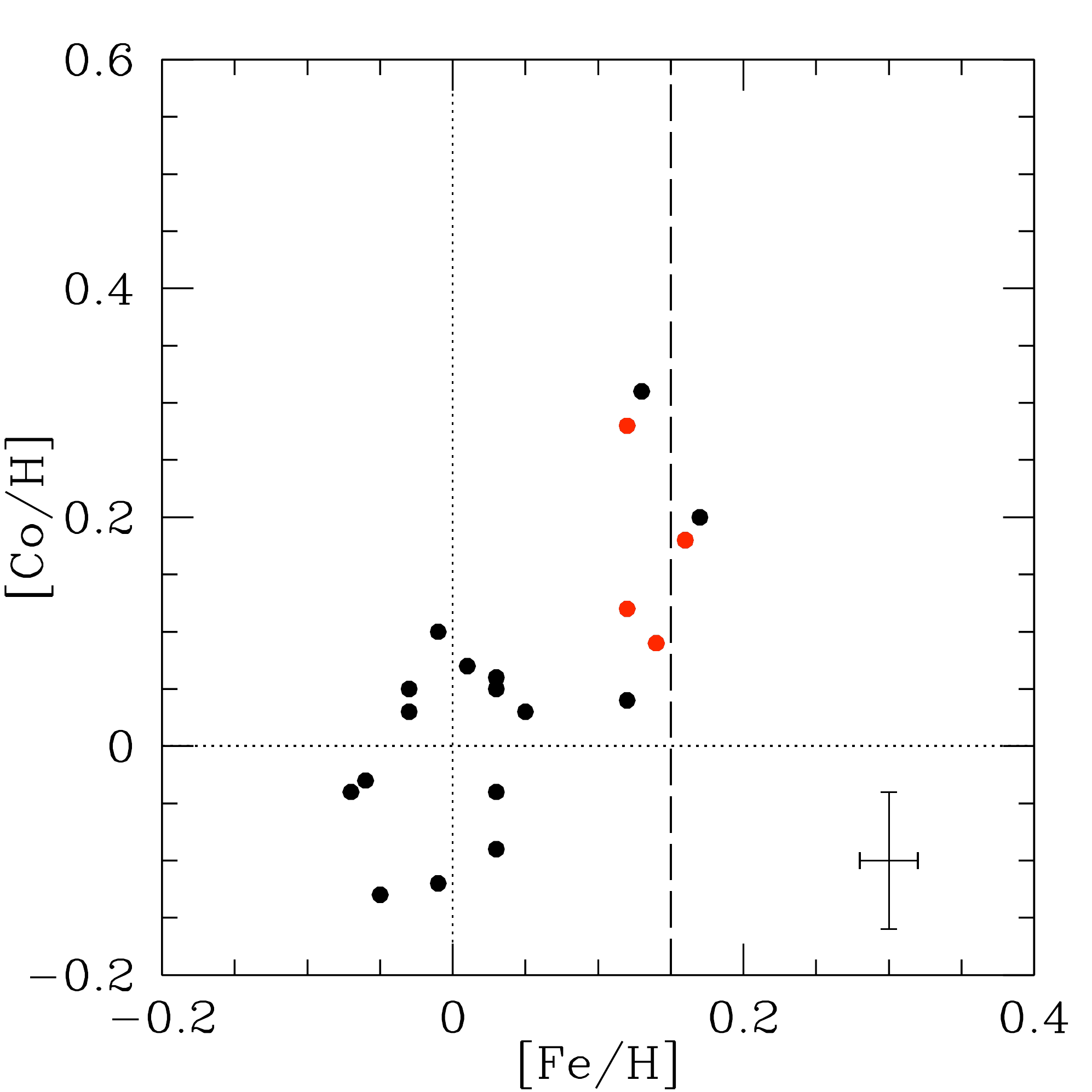}
     \caption{[X/H] vs. [Fe/H] for the sample stars (filled circles). The grey crosses are thin disc stars of \citet{bensby05}.}
        \label{abplot_coni}
  \end{figure}

\begin{figure}
  \centering
  \includegraphics[scale=0.4]{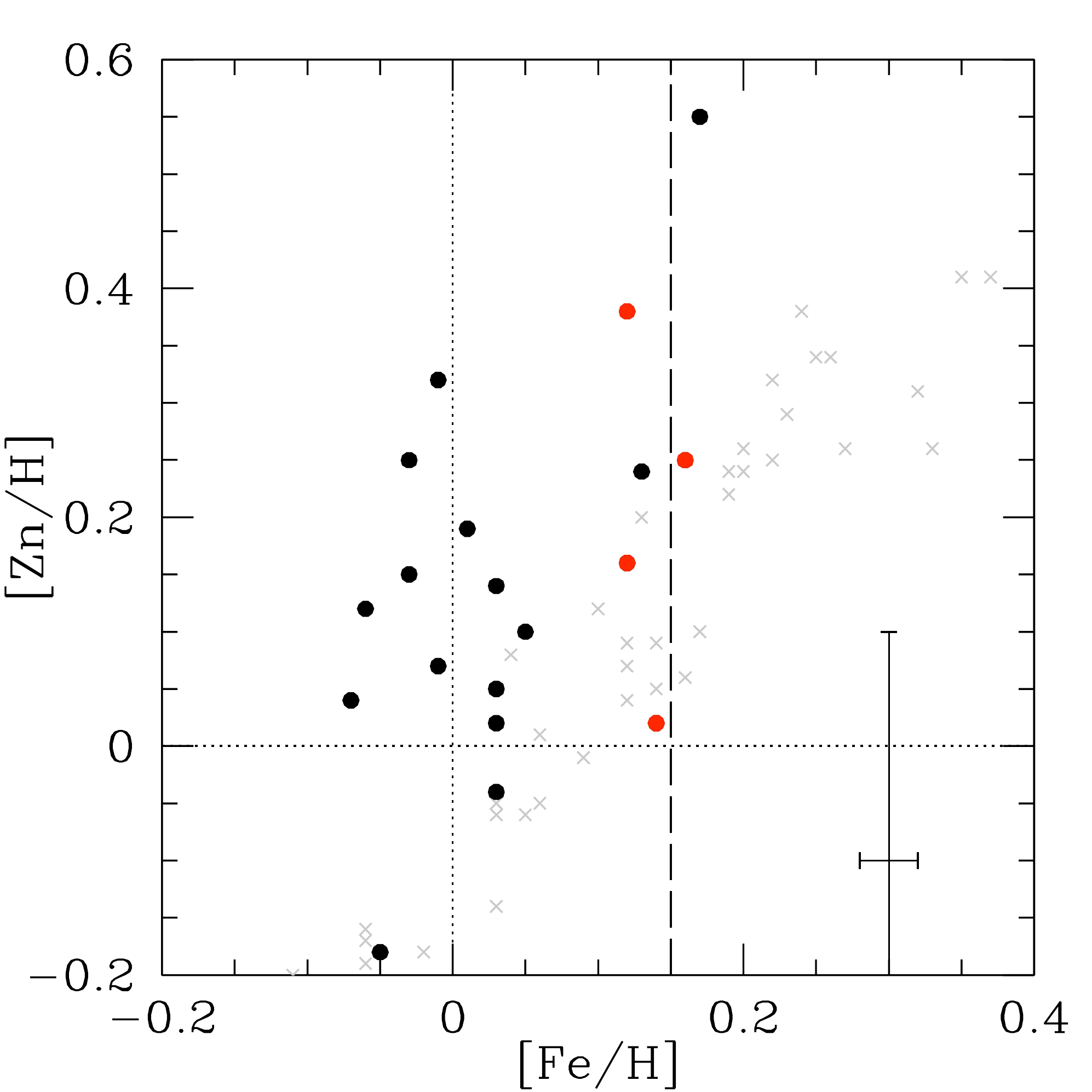}
  \includegraphics[scale=0.4]{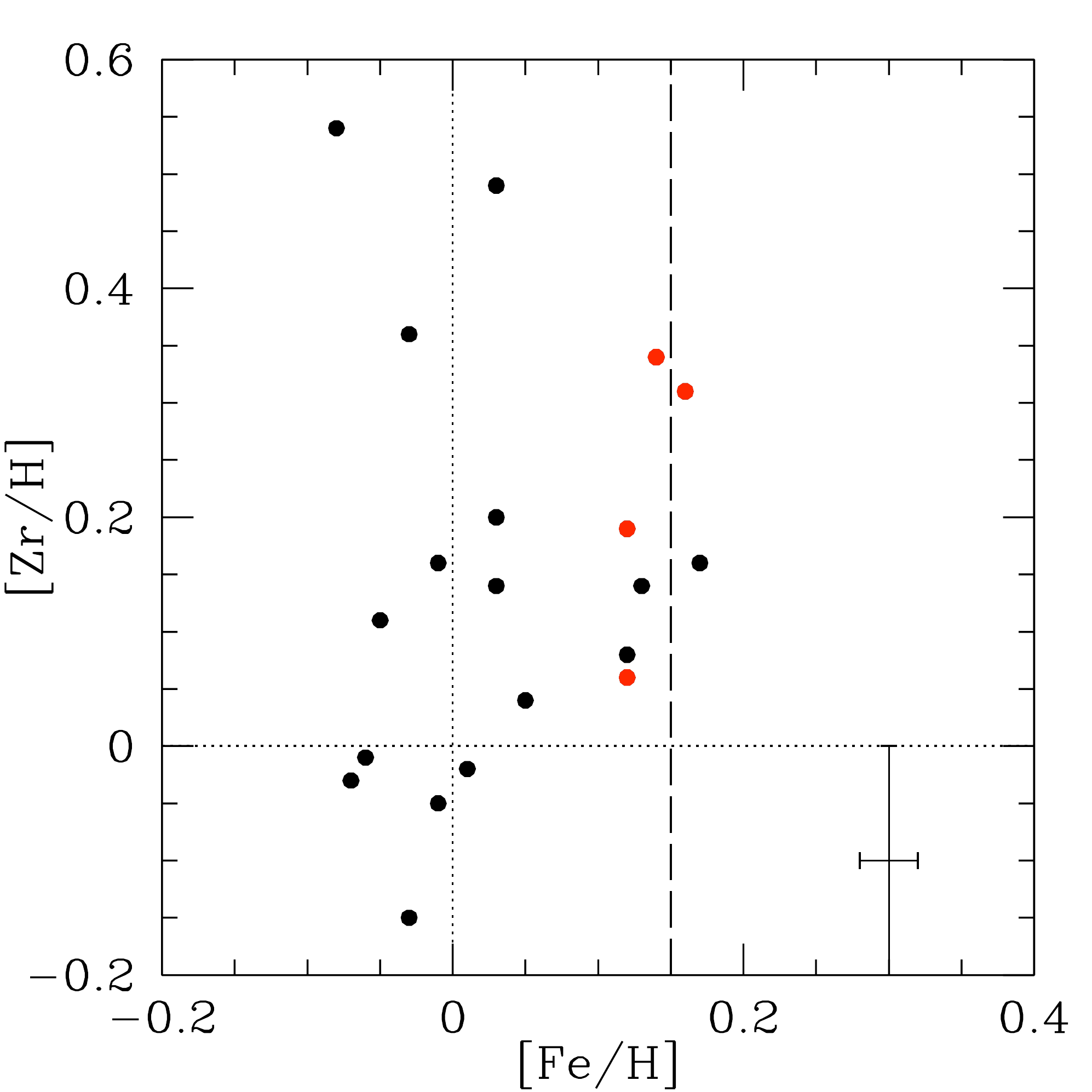}
     \caption{[X/H] vs. [Fe/H] for the sample stars (filled circles). The grey crosses are thin disc stars of \citet{bensby05}.}
        \label{abplot_znzr}
  \end{figure}

\begin{figure}
  \centering
  \includegraphics[scale=0.4]{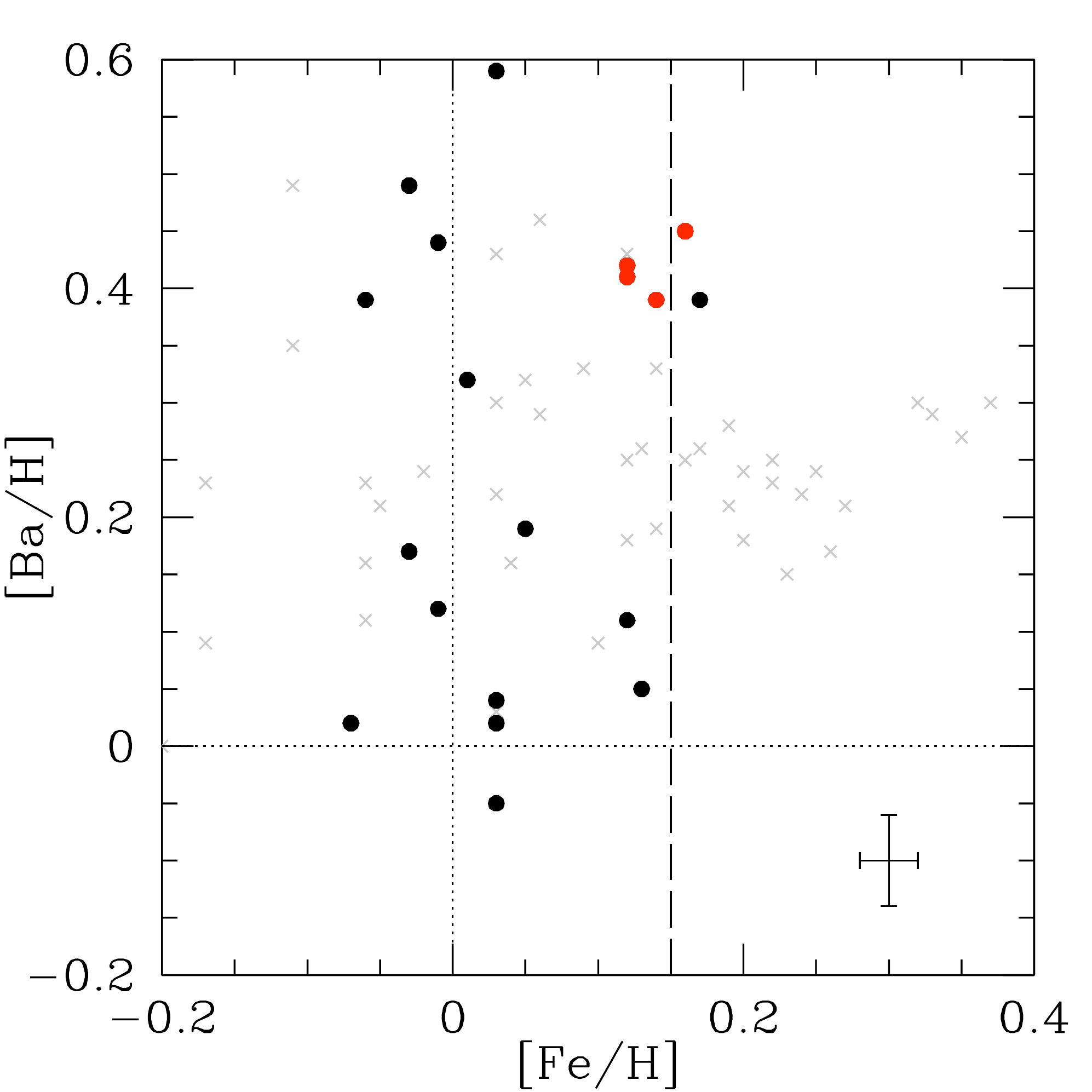}
  \includegraphics[scale=0.4]{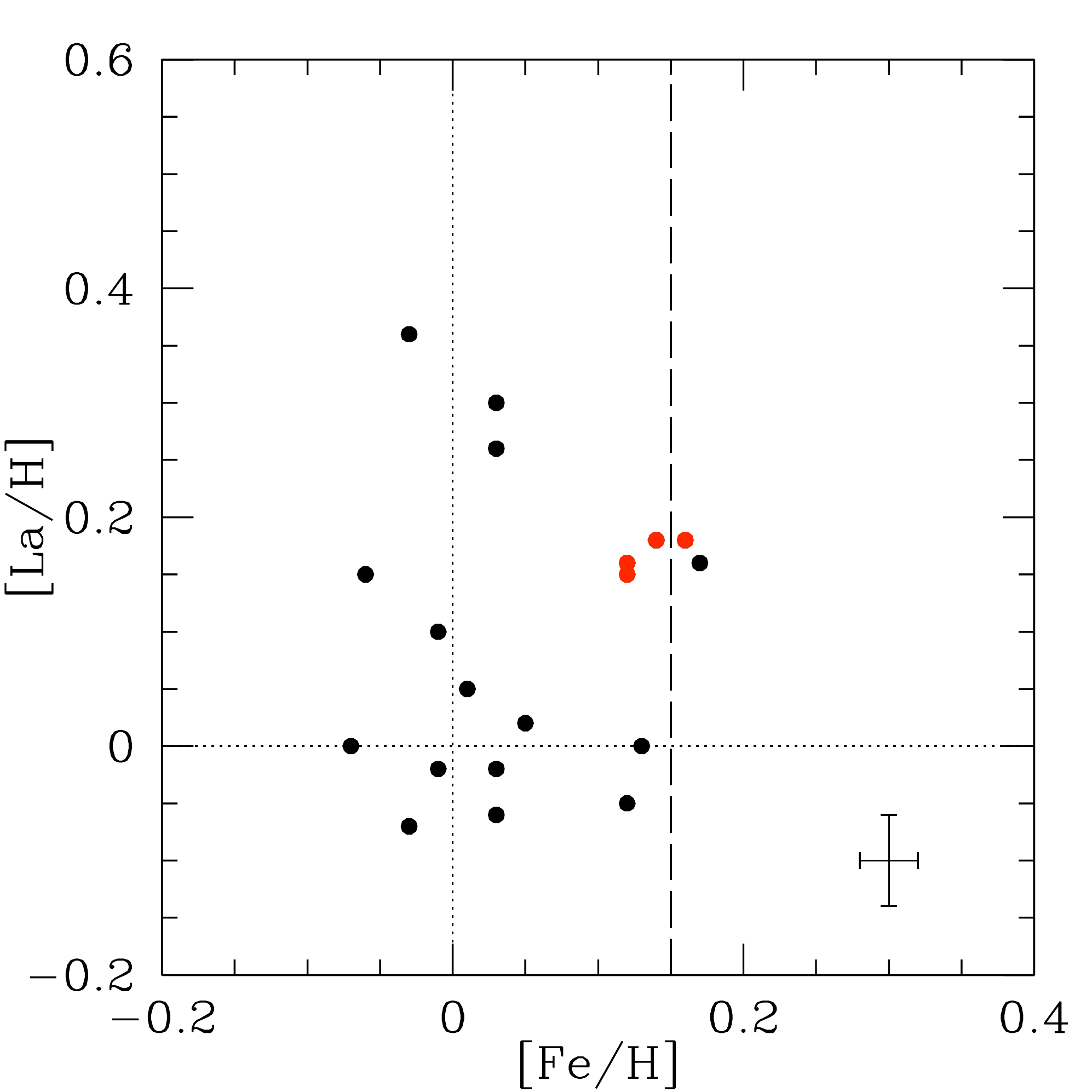}
     \caption{[X/H] vs. [Fe/H] for the sample stars (filled circles). The grey crosses are thin disc stars of \citet{bensby05}.}
        \label{abplot_bala}
  \end{figure}

\begin{figure}
  \centering
  \includegraphics[scale=0.4]{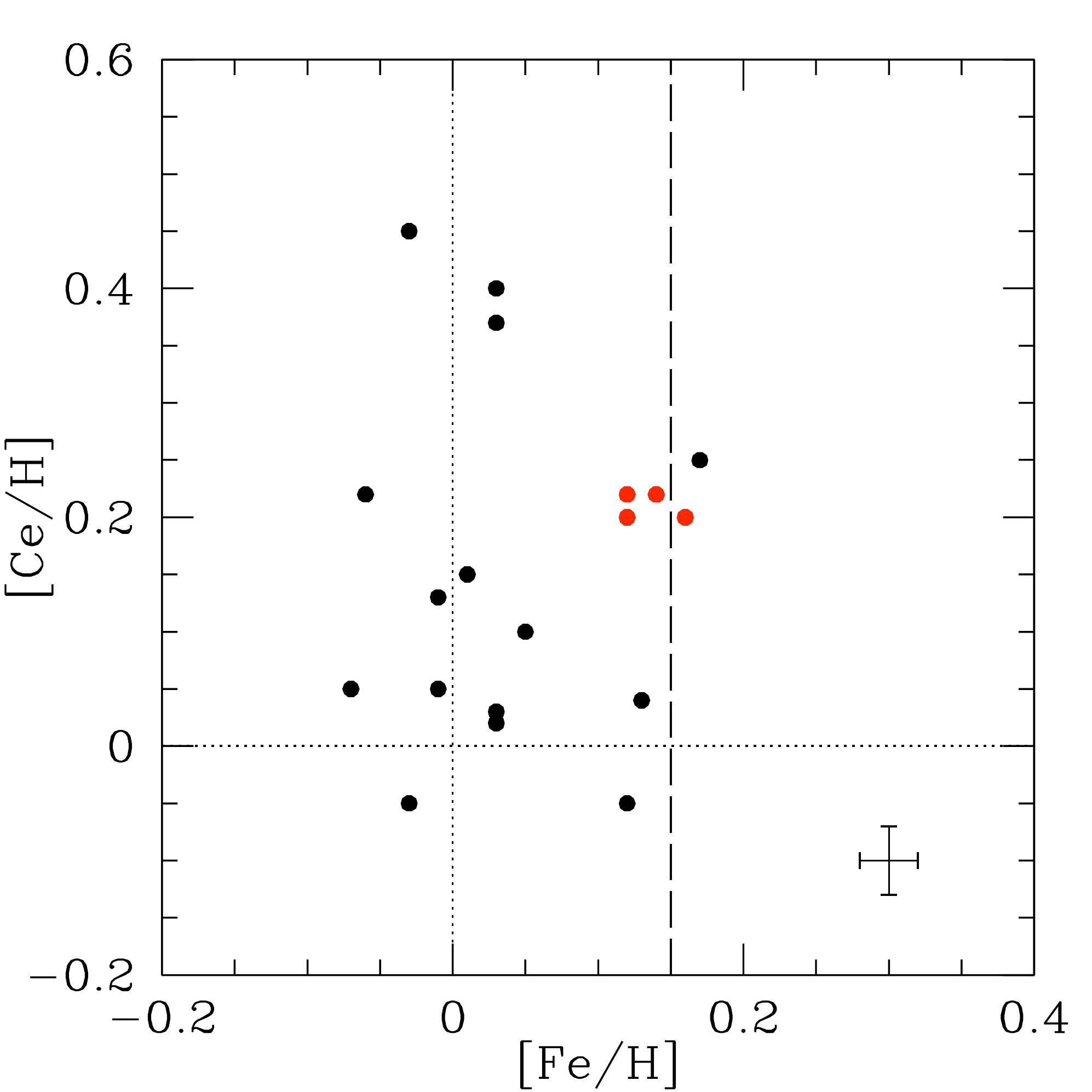}
     \caption{[X/H] vs. [Fe/H] for the sample stars (filled circles). The grey crosses are thin disc stars of \citet{bensby05}.}
        \label{abplot_ce}
  \end{figure}

Adopting the spectroscopic parameters for the remaining warmer stars, we derive the abundances of a range of elements from Na to Ce. We adopt the new solar values of \citet{asplund09} to obtain elemental abundances relative to the Sun. The final stellar abundances in [X/H] form are presented in Table \ref{ab_table}. The results are displayed graphically in Figures \ref{abplot_namg} to \ref{abplot_ce}, where we over plot the abundances of the disc field stars from \citet{bensby05} (grey crosses) where available.

\subsubsection{Errors on the derived abundances}

Possible sources of uncertainty are the errors associated with EW measurements, continuum placement and stellar parameters, as well as the number of lines used to calculate the final abundances.  The typical error in the stellar parameters were estimated to be $\delta$\teff = 50K, $\delta$log \emph{g} = 0.1 cm\,s$^{-2}$ and $\delta \xi$ = 0.1 \kms\  based on our spectroscopic derivation. The error in EWs was estimated by repeated measurements of each line during the EW measurement process. The measurement error for the synthesized elements is the uncertainty on the best fitting abundance value. Abundance dependencies on the stellar parameters and the measurement error for each element are given in Table \ref{error} for star HD78402, which has a mid range \teff and log~g.\\

S, Zn and Zr have uncertainties larger than 0.1 dex. For S and Zn, the dominant error is the large variation between the two lines studied per element, while Zr is most sensitive to temperature. No trends with \teff\  were seen for any of the elements, but it is possible that the weak lines are affected by blending, giving rise to inaccurate abundances. Na, Sc and Co have uncertainties larger than 0.05 dex, while all other element uncertainties are within the 0.05dex level. Na abundance accuracy varies with both the error in EW measurement and the line-to-line scatter. For Sc and Co, the total uncertainty is from a combination of factors given in Table \ref{error}. Note that our analysis of Sc and Co did not take into account possible effects of hyperfine structure, as we assumed such effects were negligible for our targeted lines \citep{mcwil95}. Likewise no NLTE effects were considered in this study. Such effects however may result in larger scatter and overestimated abundances. \\

\begin{landscape}
\begin{table}
\caption{Elemental abundances}             % title of Table
\label{ab_table}      % is used to refer this table in the text
\centering                          % used for centering table
\scriptsize
\begin{tabular}{ccccccccccccccccccccc}        % centered columns (4 columns)
\hline\hline                 % inserts double horizontal lines
ID &    [Fe/H]& [Na/H]& [Mg/H]& [Al/H]&[Si/H] & [S/H] &[Ca/H] &[Sc/H] & [Ti/H]& [Cr/H]& [Mn/H] & [Co/H]& [Ni/H]&[Zn/H]&[Zr/H] & [Ba/H] & [La/H] & [Ce/H]\\
\hline    
HD84598 &-0.01 &  0.17 &  0.10 &  0.05 	&  0.05 & -0.06 &  0.02 &  0.04 & -0.08 & -0.02 & -0.08	&-0.12 & -0.05 &  0.07 &  0.16  & 0.42 & 0.10 & 0.13\\
HD83234 & 0.03 &  0.11 &  0.00 &  0.31 	&  0.15 &  0.21 	&  0.01 &  0.05 &  0.20 &  0.08 & 0.03	& 0.06 &  0.07 & -0.04 &  0.49  & 0.00  & 0.26 & 0.37\\
HD81278 & 0.12 &  0.08 &  0.13 &  0.20 	&  0.27 &  0.38 	&  0.14 &  0.15 &  0.08 &  0.12 &0.12		&  0.28 &  0.21 &  0.38 &  0.06  & 0.40 & 0.16 & 0.20\\
HD80571 & 0.05 &  0.02 &  0.03 &  0.12 	&  0.17 &  0.02 	& -0.05 &  0.14 & -0.11 & -0.04 &  -0.08	&0.03 &  0.04 &  0.10 &  0.04   & 0.17 & 0.02 & 0.10\\
HD78402 &-0.03 & -0.02 &  0.15 &  0.14 	&  0.05 &  0.01 	&  0.07 &  0.17 &  0.19 &  0.07 &  -0.03	&0.03 &  0.07 &  0.15 &  0.36    & 0.47 & 0.36 & 0.45\\
HD78204 & 0.14 &  0.34 &  0.22 &  0.24 	&  0.18 &  0.17 	&  0.19 & -0.02 &  0.04 &  0.16 &  0.16	&0.09 &  0.12 &  0.02 &  0.34    & 0.37 & 0.18 & 0.22\\
HD78002 & 0.12 &  0.32 &  0.19 &  0.18 	&  0.23 &  0.33 	&  0.16 &  0.14 &  0.07 &  0.14 &  0.13	&0.12 &  0.16 &  0.16 &  0.19    &0.40&  0.15  &  0.22\\
HD77241 & 0.13 &  0.34 &  0.19 &  0.36 	&  0.41 &  0.38 	&  0.08 &  0.14 &  0.09 &  0.16 &  0.16	&0.31 &  0.24 &  0.24 &  0.14    &0.03&  0.00  &  0.04 \\
HD76128 & 0.17 &  0.11 &  0.10 &  0.25 	&  0.32 &  0.29 	&  0.03 &  0.28 &  0.05 &  0.15 &  0.12	&0.20 &  0.19 &  0.55 &  0.16    &0.37&  0.16  &  0.25 \\
HD75058 &-0.03 &  0.02 &  0.04 &  0.07 	&  0.12 &  0.02 	&  0.02 &  0.00 & -0.09 & -0.06 &  -0.05	&0.05 & -0.01 &  0.25 & -0.15   &0.15& -0.07  & -0.05\\
HD74900 &-0.01 &  0.07 &  0.14 &  0.20 	&  0.23 &  0.15 	&  0.01 &  0.11 &  0.00 & -0.02 &  -0.07	&0.10 &  0.05 &  0.32 & -0.05   &0.10& -0.02  &  0.05\\
HD74529 & 0.12 &  0.19 &  0.04 &  0.21 	&  0.22 &  0.27 	&  0.03 &  0.05 & -0.10 &  0.02 &  -0.04	&0.04 &  0.01 &  0.16 &  0.08    &0.09& -0.05  & -0.05 \\
HD74166 &-0.05 &  0.19 &  0.08 &  0.21 	&  0.02 &  0.24 	& -0.13 & -0.11 & -0.17 & -0.09 & ---		&-0.13 & -0.11 & -0.18 &  0.11  & --- & --- & --- \\
HD74165 & 0.03 &  0.11 &  0.12 &  0.28 	&  0.15 &  0.15 	& -0.01 &  0.01 & -0.01 &  0.05 &  -0.03	&0.05 &  0.00 &  0.14 &  0.14 &  0.02& -0.06  &  0.03 \\
HD73829 &-0.07 &  0.07 & -0.02 &  0.07 	&  0.15 &  0.00 	& -0.11 & -0.08 & -0.12 & -0.11 & 0.01	&-0.04 & -0.06 &  0.04 & -0.03 &  0.00&  0.00  &  0.05\\
HD73657 & 0.03 &  0.18 &  0.07 &  0.17 	&  0.25 &  0.15 	& -0.09 & -0.08 & -0.15 & -0.06 & -0.04	&-0.04 & -0.03 &  0.02 &  0.03  & -0.07& -0.02  &  0.02\\
HD72630 & 0.16 &  0.28 &  0.25 &  0.28 	&  0.24 &  0.19 	&  0.19 &  0.16 &  0.11 &  0.19 & 0.17	& 0.18 &  0.17 &  0.25 &  0.31  &  0.43&  0.18  &  0.20\\
HD72320 & 0.03 &  0.16 &  0.07 &  0.12 	&  0.09 &  0.05 	&  0.08 &  0.04 &  0.00 &  0.05 & 0.01	&-0.09 & -0.02 &  0.05 &  0.20 &  0.57&  0.30  &  0.40 \\
HD69836 & 0.01 & -0.01 &  0.09 &  0.12 	&  0.17 &  0.06 	& -0.03 &  0.11 & -0.09 & -0.04 & -0.01	& 0.07 &  0.07 &  0.19 & -0.02 &  0.30&  0.05  &  0.15\\
HD122721 &-0.06 &  0.00 &  0.01 &  0.08 &  0.11 &  0.00 	& -0.04 &  0.02 & -0.12 & -0.07 & -0.04	&-0.03 & -0.02 &  0.12 & -0.01&  0.37&  0.15  &  0.22\\
\hline
\end{tabular}
\end{table}
\end{landscape}

\begin{landscape}
\begin{table}
%\begin{sidewaystable*}
\caption{Abundance Sensitivities for star HD78402}             % title of Table
\label{error}      % is used to refer this table in the text
\centering
\footnotesize
\begin{tabular}{lccccccccccccccccccccccccc}        % centered columns (4 columns)
\hline\hline                 % inserts double horizontal lines
Parameter&Fe & Na & Mg & Al & Si & S & Ca & Sc & Ti &Cr & Mn & Co & Ni& Zn & Zr & Ba & La & Ce\\
\hline    

T$_{\rm eff}\pm$50	& $\pm$0.01 & $\pm$0.04 & $\pm$0.03 & $\mp$0.02 & $\pm$0.03 & $\pm$0.01 & $\pm$0.02 & $\mp$0.01 & $\pm$0.01 & $\pm$0.02 &  $\pm$0.01 & $\pm$0.01 & $\pm$0.01& $\pm$0.05 & $\pm$0.08 &$\pm$0.02 & $\pm$0.02 &$\pm$0.02 \\
log $g \pm$0.1		&$\pm$0.01  & $\mp$0.02 & $\mp$0.04 & $\pm$0.01 & $\mp$0.04 & $\mp$0.01 & $\mp$0.01 & $\pm$0.01 & $\pm$0.04 &$\pm$0.02  & 	0.0	   &$\pm$0.02  & $\pm$0.03& $\pm$0.01 & $\pm$0.02 &$\pm$0.01 &$\pm$0.03  &       	0.00\\
$\xi\pm$0.1 		&$\mp$0.02  & $\mp$0.02 & $\mp$0.02 & $\mp$0.01 & $\mp$0.04 & $\pm$0.02 & $\mp$0.04 & $\mp$0.03 & $\mp$0.05 & $\pm$0.01 & $\mp$0.01  & $\pm$0.02 & 0.00 	    & $\pm$0.01 & $\pm$0.02 & $\mp$0.03& $\mp$0.02&  		0.00\\
$\Delta$ EW / Synth &$\pm$0.02  & $\pm$0.06 & $\pm$0.05 & $\pm$0.04 & $\pm$0.03 & $\pm$0.03 & $\pm$0.03 & $\pm$0.03 & $\pm$0.02 & $\pm$0.02 & $\pm$0.02  & $\pm$0.02 & $\pm$0.02 & $\pm$0.03 & $\pm$0.03 & $\pm$0.02&$\pm$0.02 &$\pm$0.02 \\
$\sigma$(line-to-line)& $\pm$0.01& $\pm$0.07& $\pm$0.03 & ---                 & $\pm$0.02 & $\pm$0.1    & $\pm$0.02 & $\pm$0.06 & $\pm$0.03 & $\pm$0.01& $\pm$0.01  & $\pm$0.05 & $\pm$0.02 & $\pm$0.2   & $\pm$0.03 & $\pm$0.04 & ---              &          --- \\  
\hline
Total 			&$\pm$0.02 & $\pm$0.09 & $\pm$0.05 & $\pm$0.04 & $\pm$0.05 & $\pm$0.11 & $\pm$0.03 & $\pm$0.06 & $\pm$0.03 & $\pm$0.04 & $\pm$0.02 &$\pm$0.06 & $\pm$0.05 & $\pm$0.2 & $\pm$0.1 & $\pm$0.04&$\pm$0.04& $\pm$0.03 \\
\hline
\end{tabular}
%\end{sidewaystable*}
\end{table}
\end{landscape}

\section{Origin of the Hyades Supercluster}

\subsection{Chemical tagging}

In the case of a dispersing stellar aggregate, where the member stars were all formed from a common proto-cluster gas cloud, we expect to find chemical homogeneity \citep{desilva06,desilva07}. The abundance results presented in section 4.2 show the abundance distribution clustered around [Fe/H] = 0.0 and [Fe/H] = 0.15. From Fig \ref{fe_teff} we see the majority of the stars are located around solar metallicity, while 7 stars have high metallicities comparable to the Hyades open cluster metallicity. Note that the [Fe/H] distribution is such that there is a clear gap between 0.05 $<$ [Fe/H] $<$ 0.1, rather than a heterogeneous spread across the sample. The individual element abundance plots, show that four of these 7 metal-rich stars (HD81278, HD78204, HD78002, HD72630) are clumped together across many of the elements. The standard deviation in abundance for these four stars is less than 0.05 for all elements except for Na, S, Sc, Co, Zn and Zr.  Note that these elements also have a higher abundance uncertainty, whereas all other elements were measured to 0.05 dex internal accuracy.\\

We compare our sample star abundances against the thin disc abundances of \citet{bensby05}. The majority of our sample contains solar-level metallicity stars, which match the field star abundances well, implying that our sample of potential Hyades supercluster is dominated by field stars, rather than those from a dispersing stellar aggregate. This supports the case of the kinematically defined Hyades supercluster being a transient feature caused by disc dynamics. Regarding those stars at [Fe/H] $\sim$ 0.15, the disc field also contain similar high metallicity stars which overlap with the Hyades open cluster abundances. We ask how likely is it that the higher metallicity stars in our sample are Hyades-related, and not just metal-rich field stars? \\

Firstly Na is enhanced relative to the field stars. This is seen for some solar-metallicity level stars as well as few of the metal-rich stars. A possible explanation for this is the abundance variations between giants and dwarfs \citep{ona}. It is possible that internal mixing mechanism have enhanced the natal Na abundances in some of these giant stars, hence making the measured abundances less reliable for chemical tagging.  More interestingly Ba is much enhanced among the four metal-rich stars relative to the field star abundances. A similar Ba enhancement is seen in the Hyades open cluster based on high resolution abundance analysis \citep{desilva06}. In fact the observed enhancement in these four stars for La and Ce is also seen in the Hyades open cluster. Using the results of \citet{desilva06}, the Hyades open cluster mean abundances are [Ba/H] = 0.45, [La/H] = 0.10 and [Ce/H] = 0.17, when scaled to the solar-values of \citet{asplund09} which we employ in this analysis.  The average abundance values of the four metal-rich stars in this study are [Ba/H] = 0.40, [La/H] = 0.16 and [Ce/H] = 0.2. This is in remarkable agreement with the Hyades open cluster abundances taking into account the measurement uncertainties in the two studies. We also attempted to compare our results with those by \citet{paulson} for the Hyades open cluster for alpha-element and Fe-peak element abundances. However upon investigation, we were unable to determine their absolute solar-abundance scale to enable a useful comparison. Since their work was focused on differential abundances relative to their measured solar spectrum, many unquantifiable systematics likely exist between the two studies.\\

The combination of Hyades open cluster-like abundance patterns for Fe and heavy elements and the high level of abundance homogeneity seen among the four metal-rich stars of our sample, suggest that these four stars were once part of the Hyades star-forming aggregates, but are now dispersed elsewhere. Combining this scenario with the observed spread in abundances for the majority of our sample stars adds further mystery to the origin of the Hyades supercluster. Therefore we now turn to explore the kinematical membership criteria.\\

While open cluster members are easily determined from radial velocities, memberships for dispersed groups are based on the star's UVW space motion. However the exact criteria for selecting members remain somewhat uncertain. Eggen's method of isolating dispersed  members required member stars to have the same V-velocity, i.e. in the direction of the Galactic rotation, based on epicycle theory. However, recent more realistic Galaxy dynamical models show that such a clump in V is not necessary for membership to a moving group or supercluster; rather a small $tilt$ or $curve$ in the distribution in the U-V plane is the dynamical signature of a dispersed aggregate \citep{skul,fh,helmi,hr1614}. Along with our abundance results, we explore the membership probabilities of our sample stars using the positions on the color-magnitude diagram and the stellar kinematics.\\

\subsection{Distances and Color-Magnitude diagram}
\begin{figure}
  \centering
  \includegraphics[scale=0.4]{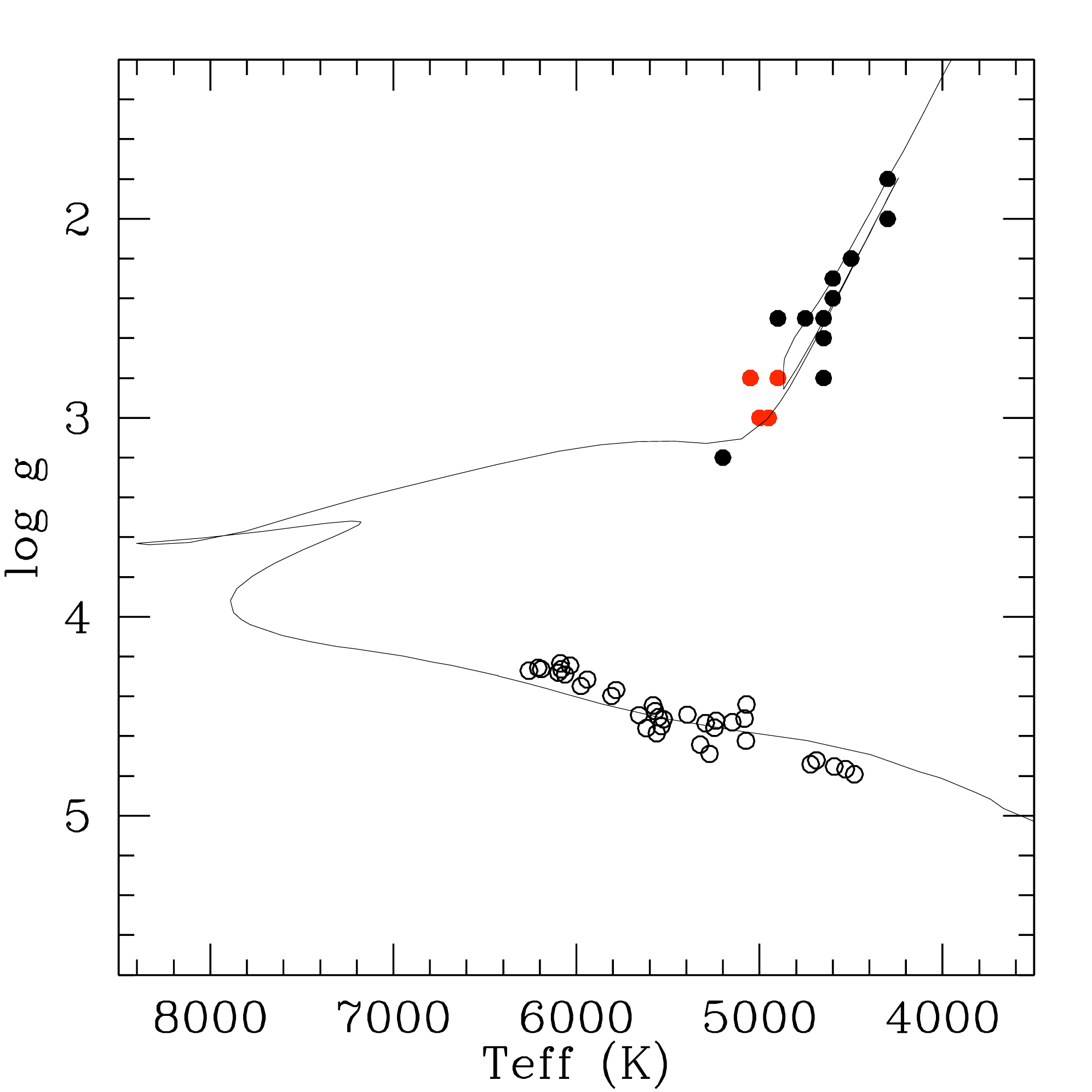}
     \caption{Spectroscopic \teff vs. log g of the supercluster sample and the Hyades open cluster dwarfs (open circles). The red circles are the four metal-rich stars discussed in section 5.1.}
        \label{cmd_teff}
  \end{figure}

A fundamental property of a star cluster, as a single-generation population, is that it displays a clear evolutionary sequence of stars of various masses in the CMD, contrary to the field star population which contains a large range in age and metallicity. We compare our sample stars against the Hyades open cluster CMD. Since only 10 stars in our sample have accurate distance measurements in the literature, we plot the spectroscopically derived Teff vs. log $g$ of the studied sample. Fig \ref{cmd_teff} shows our sample together with the spectroscopically derived stellar parameters of the Hyades open cluster dwarfs (open circles) by \citet{primas}. We can confidently make this comparison because the \teff\ and log $g$ for both open cluster dwarfs and the supercluster giants were derived via the same methodology for similar quality UVES spectra, using the same Fe {\sc i} and Fe {\sc ii} lines and models atmospheres. To guide the eye, overlaid in Fig \ref{cmd_teff} is an isochrone of 0.7 Gyr, Z = 0.025 from \citet{girardi00}, which closely corresponds to the Hyades open cluster age and metallicity. The larger red circles are the four Hyades open cluster-like stars discussed above in section 5.1.\\

Figure \ref{cmd_teff} shows that all of our sample stars are consistent with a Hyades-aged isochrone. However our chemical abundance results in section 5.1 show only four stars are likely associated with the Hyades star-forming event. Therefore lying along the isochrone alone is not a sufficient condition for group membership.\\

\begin{figure}
  \centering
  \includegraphics[scale=0.4]{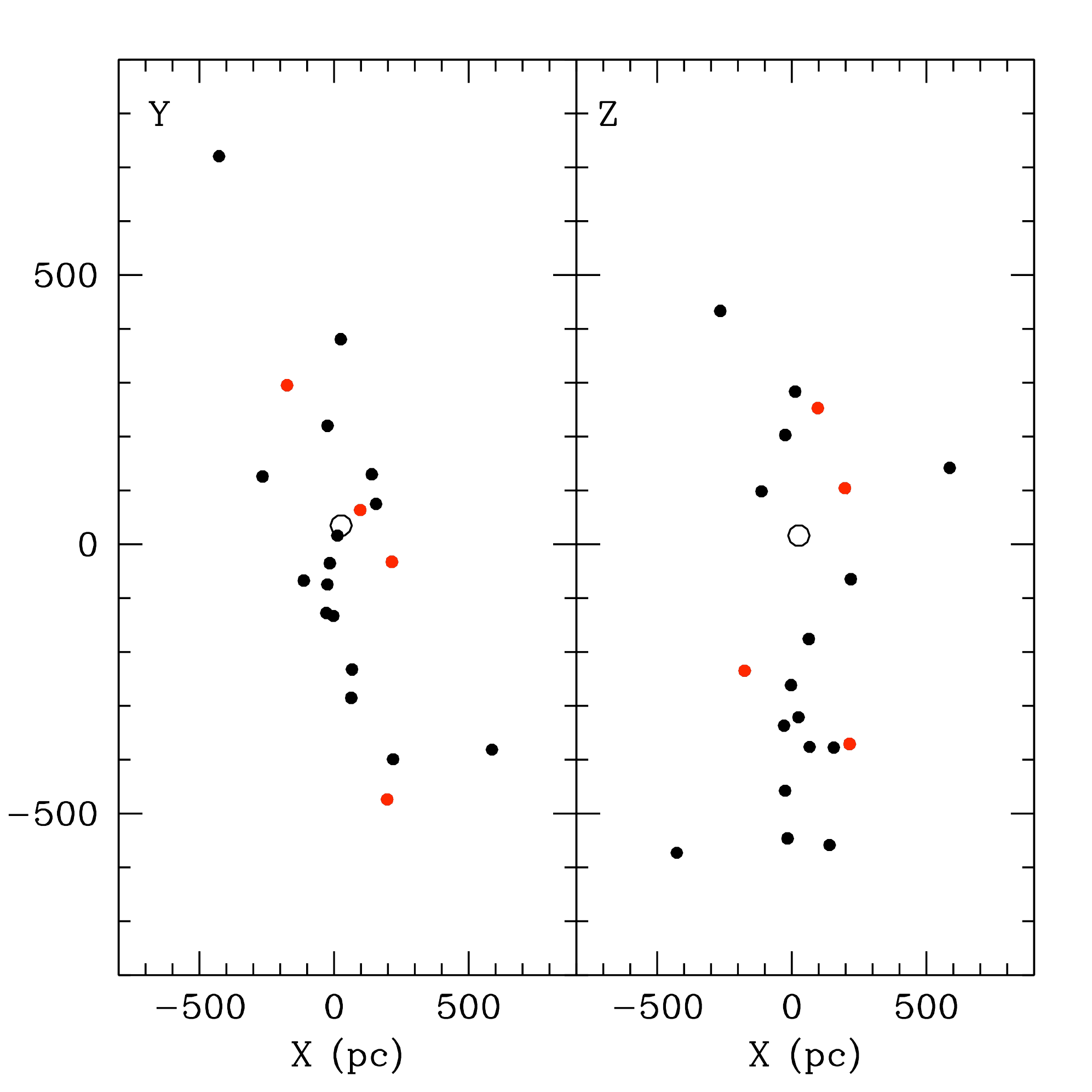}
     \caption{The location of the sample stars in X-Y and X-Z plane, where X-Y is in the Galactic plane. The open circle represents the Hyades open cluster and the red stars are the four metal-rich stars with Hyades-like abundances discussed in section 5.1.}
        \label{xyz}
  \end{figure}

Using Figure \ref{cmd_teff} we can use the best fitting isochrone to derive the distances for each of our sample stars. By comparing the location of each star on the isochrone we determined the equivalent absolute V magnitude from the isochrone. With the determined absolute magnitudes and the known apparent magnitudes we derive the distances of the sample stars. Table \ref{distances} lists the derived distances and Figure \ref{xyz} shows the location of our stars in X-Y and X-Z spatial co-ordinates, where X-Y is in the Galactic plane. The open circle represents the location of the Hyades open cluster. The four red stars with Hyades-like abundances lie outside the open cluster and are not present members of the bound open cluster.\\

\begin{table*}
\caption{Absolute magnitudes and Distances}             % title of Table
\label{distances}      % is used to refer this table in the text
\centering                          % used for centering table
\begin{tabular}{c c c c c c}        % centered columns (4 columns)
\hline\hline                 % inserts double horizontal lines
Star ID &  M(V) & Distance (pc) \\    % table heading 
\hline   
HD84598 &   1.355 &   164.8   \\
HD83234 &   1.162 &    300.3   \\
HD81278 &   1.250 &   278    \\
HD80571 &   0.612& 284.2   \\
HD78402 &   0.612 &  713.8 \\  
HD78204 &   1.345 &  523.6  \\
HD78002  &  1.345 &  415.9  \\ 
HD77241 &  -0.052 & 1015   \\  
HD76128  & -0.369 & 498.7  \\
HD75058  &  0.686 & 447.5  \\ 
HD74900 &   0.620 & 293.8  \\ 
HD74529 &   0.620 &  361.4 \\ 
HD74166 &  -0.725 & 464.5  \\ 
HD74165 &   0.417 & 547.8  \\ 
HD73829  &  0.555 &  590.2 \\ 
HD73657 &  -0.252 &  415.3 \\ 
HD72630 &   1.313 & 429.9  \\  
HD72320 &   1.285 & 341.2  \\ 
HD69836 &   0.246 & 460  \\ 
HD122721 &  0.555&  523.6 \\ 
\hline                                   %inserts single line
\end{tabular}
\end{table*}

\subsection{Radial velocities}

Radial velocities for our target stars were calculated from our high-resolution spectra using the DAOSPEC code \citep{stetson}, which matches lines with measured equivalent widths against a line list. A comprehensive and realistic line list for stellar spectra was obtained using the \textit{extract stellar} on the VALD database. We used our line list for a [Fe/H] $=-0.7$, $T_\textrm{eff}=5500$, $\log g=4.25$, $\xi=1.3$ star, where for our radial velocity calculation of Hyades stars we can expect little effect of the stellar parameter differences. DAOSPEC was run on three $100$ \AA\  intervals at $5600-5700$, $5800-5900$ and $6000-6100$ \AA\, where DAOSPEC returns a radial velocity estimate with a corresponding dispersion by comparison with the line list. The radial velocity was obtained for each star from a weighted average of the three measurements, after correcting to heliocentric radial velocity. The final values are listed in column 2 of Table \ref{velocities}.

\subsection{U,V,W velocities}
\begin{figure}
  \centering
  \includegraphics[scale=0.4]{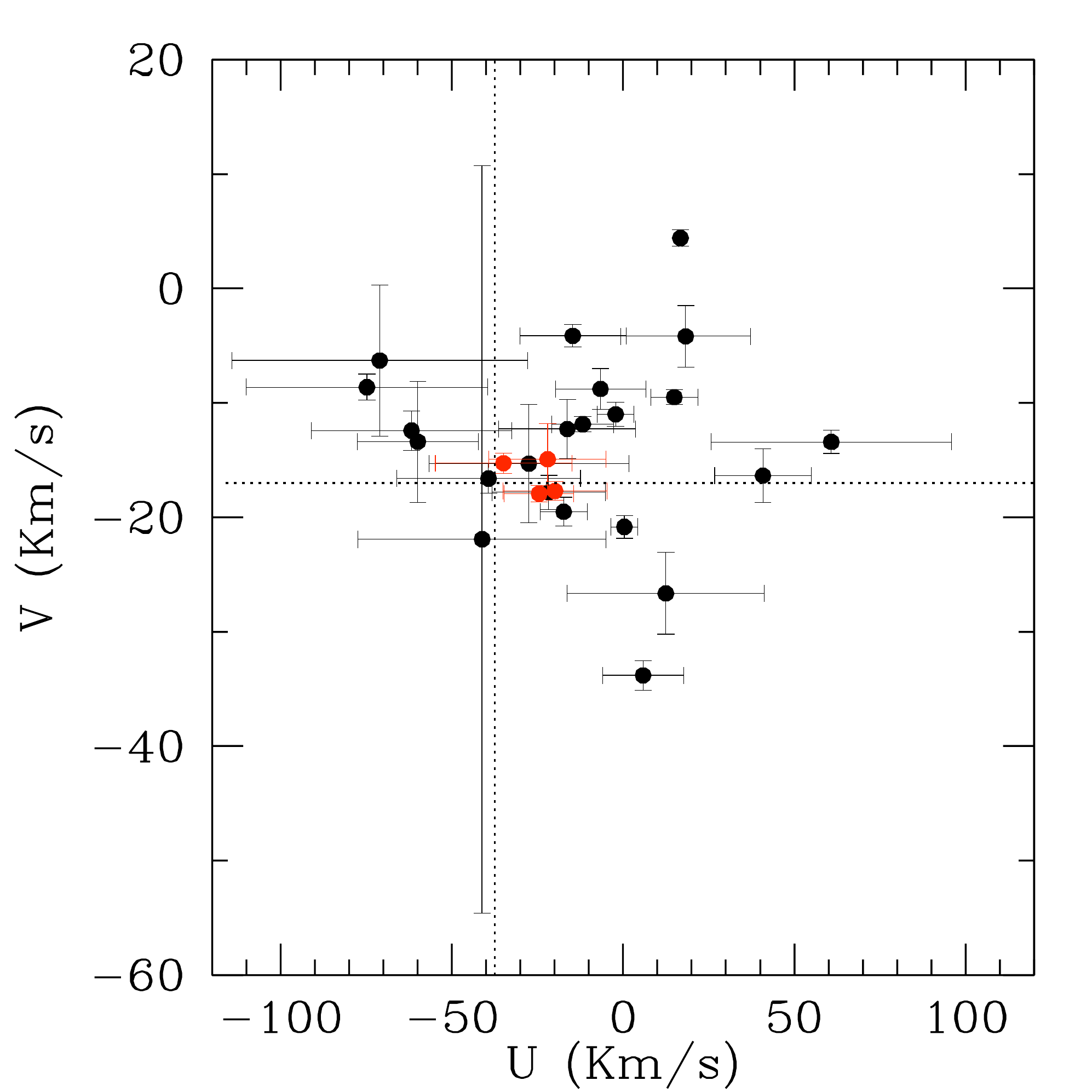}
     \caption{Galactic V velocity vs. U velocity for the sample stars. Red stars are the four metal-rich stars discussed in section 5.1. The dotted lines mark the Hyades open cluster velocities.}
        \label{uvw}
  \end{figure}

Using the derived distances, radial velocities, and proper motions from the Tycho-2 catalogue we calculate the space velocities assuming a right handed Cartesian $UVW$ velocities centred on the solar position, with the $U$ velocity positive towards the Galactic centre. Velocities were corrected to the Local Standard of Rest (LSR) using the peculiar motion of the sun given by \citet{dehnen98} of $10,\ 5.3,\ 7.2 \ \kms$ in $U$, $V$ and $W$ respectively. Errors in $UVW$ were calculated using a Monte-Carlo simulation of 1000 realizations to propagate the errors in distance (where we assumed a $20\%$ error), proper motion and radial velocity into the space velocities, from whence the dispersions were calculated. The results are tabulated in Table \ref{velocities} and the V vs. U values are plotted in Figure \ref{uvw}, where the dotted line marks the V velocity of the Hyades open cluster and the red circles highlight the four Hyades open cluster-like stars discussed earlier in section 5.1.\\

\begin{figure}
  \centering
  \includegraphics[scale=0.4]{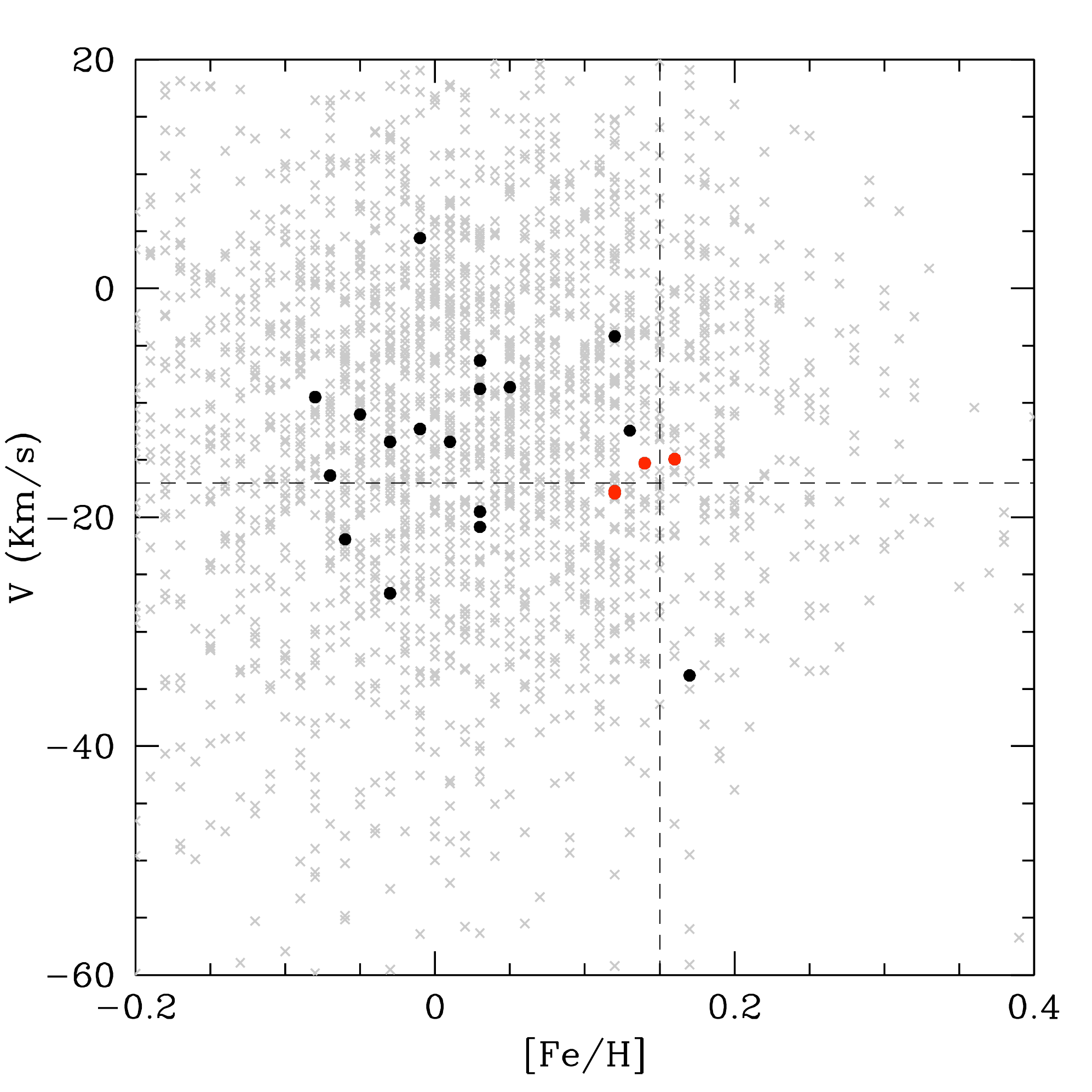}
     \caption{Galactic V velocity vs. [Fe/H] for the sample stars. Red stars are the four metal-rich stars discussed in section 5.1. The grey crosses are the output from the Besancon stellar population model for stars based on our selection criteria.}
        \label{bes_mod}
  \end{figure}

We find that our sample stars are located within $\pm$ 20 \kms of the Hyades cluster V velocity. Stars with large velocity deviations are most likely unrelated field stars based on the kinematical membership criterion. This shows that the hypothesis used by Wilson to select probable Hyades supercluster members was not efficient. The four metal-rich stars with Hyades open cluster-like abundances lie tightly clustered within 3 \kms\ of the Hyades cluster velocity, along with several other stars which did not have similar elemental abundances. To estimate the expected number of background stars with Hyades-like motion and metallicity, we queried the Besancon stellar population model \citep{besancon} based on our sample selection criteria. The model was queried for stars within 600pc, with magnitude V $<$ 10, between galactic longitudes 253 $< l <$ 279 and latitudes -3 $< b <$ 3. The [Fe/H] versus V velocity of the model stars, as well as our sample stars are plotted in Figure \ref{bes_mod}. The red circles are the four stars with Hyades open cluster-like abundances. Two of those stars have almost identical [Fe/H] values and V velocities, hence they are overlapping in Figure \ref{bes_mod}. Another star, HD77241, also lies close the open cluster velocity. Despite having a similar high metallicity, this star did not have similar abundance levels across the other studied elements. Enforcing both the kinematical and chemical tagging criteria, we confirm that these four stars are likely to be former members of the Hyades open cluster.\\

%\begin{landscape}
\begin{table*}
\caption{Radial and Galactic space velocities}             % title of Table
\label{velocities}      % is used to refer this table in the text
\centering                          % used for centering table
\begin{tabular}{cccccccccc}        % centered columns (4 columns)
\hline\hline                 % inserts double horizontal lines
Star ID &  RV (\kms)& RV error& U (\kms) & U error & V (\kms) & V error & W (\kms) & W error \\    % table heading 
\hline   
HD 84598 &  1.07 & 0.71 &  15.05 &  0.51 &   4.41 &  0.72 &   1.81 &  0.59 \\
HD 83234 & 11.42 & 0.51 &  -8.68 & 13.42 &  -8.78 &  1.77 &  16.02 &  9.15 \\
HD 81278 & 20.18 & 0.54 & -26.87 & 10.28 & -17.93 &  0.72 &  26.15 & 10.06 \\
HD 80571 & 15.13 & 0.48 & -77.95 & 35.93 &  -8.63 &  1.14 &  25.25 & 35.33 \\
HD 78402 & 18.12 & 0.54 &  59.80 & 35.68 & -13.41 &  1.02 &  32.01 & 39.13 \\
HD 78204 & 19.51 & 0.68 & -37.43 & 20.28 & -15.27 &  0.89 &  10.43 & 21.79 \\
HD 78002 & 25.39 & 0.61 & -22.19 & 15.41 & -17.71 &  0.83 &   5.41 & 17.17 \\
HD 77241 & 18.74 & 0.58 & -64.74 & 29.73 & -12.43 &  1.73 &  17.54 & 33.33 \\
HD 76128 & 40.41 & 0.57 &   3.98 & 11.96 & -33.82 &  1.28 &  32.03 & 18.06 \\
HD 75058 & 27.90 & 0.49 &  10.71 & 29.27 & -26.65 &  3.56 & -54.89 & 36.91 \\
HD 74900 & 21.12 & 0.48 & -18.49 & 20.31 & -12.29 &  2.58 & -13.73 & 25.69 \\
HD 74529 & 11.50 & 0.55 &  16.59 & 19.29 &  -4.19 &  2.71 & -39.80 & 24.88 \\
HD 74166 & 18.15 & 0.66 &  -4.18 &  5.47 & -11.01 &  1.06 &   6.11 &  6.61 \\
HD 74165 & 24.77 & 0.59 & -74.18 & 43.84 &  -6.30 &  6.59 & -26.33 & 57.07 \\
HD 73829 & 17.43 & 0.53 &  39.52 & 14.36 & -16.36 &  2.34 &  11.66 & 18.65 \\
HD 73657 & 28.84 & 0.59 & -19.58 &  6.99 & -19.51 &  1.25 &  11.73 &  9.31 \\
HD 72630 & 25.62 & 0.60 & -24.37 & 17.40 & -14.92 &  3.08 &  13.99 & 11.47 \\
HD 72320 & 27.73 & 0.61 &  -1.57 &  3.95 & -20.85 &  0.99 &   3.87 &  5.54 \\
HD 69836 & 38.72 & 0.64 & -62.88 & 17.95 & -13.40 &  5.28 &  10.21 & 27.25 \\
HD 122721 & -28.80 & 0.47 & -43.79 & 36.83 & -21.92 & 32.68 &  -9.04 & 14.38 \\
\hline                                   %inserts single line
\end{tabular}
\end{table*}
%\end{landscape}

\section{Conclusion}

We obtained high resolution spectra of 45 proposed Hyades supercluster members, which were selected based on early estimates of the targets' metallicity and position along the expected dispersion orbit of the Hyades supercluster, as well as F-type Hipparcos stars from \citet{e98}. The observing results showed that only 26 stars were suitable for detailed abundance analysis. The abundance distribution showed a double-peaked distribution, with two clear clustering of stars around solar metallicity and around Hyades open cluster metallicity. \\

Four of the stars with Hyades open cluster metallicity also shared similar abundance patterns, with the star-to-star scatter within 0.05 across all elements except for Na, S, Sc, Co, Zn and Zr. Possible deviations due to stellar evolutionary effects as well as larger abundance uncertainty in this analysis make these six elements less reliable for chemical tagging than the other elements analysed in this study. Further, the four stars' abundances for Ba, La and Ce match the abundances observed in the Hyades open cluster. These heavy elements are produced predominately via the slow neutron capture process, where the likely sites of formation are Type II supernova explosion. Therefore the heavy element abundances are not modified during the lifetime of the giant and sub-giant stars we have studied, and the measured abundances should represent the abundance levels laid out during the stars' birth in a proto-cluster cloud.\\

Examining the space velocities of our targets revealed that our sample stars are within $\pm$ 20 \kms\ of the Hyades open cluster V velocity. Those velocity outliers are most likely to be non-members of the kinematically defined Hyades supercluster. Applying both kinematical and chemical criteria, we find four stars are likely dispersed members from an earlier Hyades star-forming event. These four stars share Hyades-like kinematics and chemical abundances, but spatially they are located away from the Hyades open cluster. Therefore these four stars are clearly supercluster members, rather than open cluster members. The rest of the stars are either non-members of the supercluster, or they are kinematically members of a co-moving group, but are otherwise unrelated to the Hyades star-forming event. \\

The presented results are primarily a test of the Hyades cluster disruption hypothesis of \citet{wilson}. We find $\sim$15\% of his sample are supercluster members based on kinematics and chemical abundances. Our study only included one slow rotating star from \citet{e98}, hence we are unable to test the validity of his sample. From our results, we conclude that the kinematically defined Hyades supercluster is made up at least partly of dispersed Hyades cluster stars. Further study of the Hyades supercluster, especially targeting kinematically defined dwarf stellar members is encouraged.

%\begin{longtable}
%\caption{Line list}\label{lines}
%\centering
%\begin{tabular}{c c c c c}
%Wavelength (\AA) & Species & LEP(eV) & log $gf$ \\
%\hline

\section*{Acknowledgments} %If needed

GDS would like to thank F. Primas for the use of the stellar parameters for the Hyades open cluster dwarfs based on  a  new Hyades open cluster abundance analysis in preparation.

\end{document}